\documentclass[%
 reprint,
 superscriptaddress,
 numerical,
 showpacs,
 amsmath,amssymb,
 aps,
 longbibliography,
 prx,
 floatfix
]{revtex4-1}
\usepackage{amsmath}
\usepackage{amssymb}
\usepackage{soul}
\usepackage{braket}
\usepackage{dsfont}
\usepackage{float}
\usepackage{xcolor}

\usepackage[colorlinks, linkcolor=blue,citecolor=blue, urlcolor=blue]{hyperref}
\usepackage{graphicx}

\newcommand{\tr}{\mathrm{Tr}}

\newcommand{\figref}[1]{Fig.~\ref{#1}}

\begin{document}

\title{Power-law entanglement growth from typical product states}

\author{Tal{\'i}a L.~M. Lezama}
\affiliation{Max-Planck-Institut f\"ur Physik komplexer Systeme, 01187 Dresden, Germany}
\author{David J. Luitz}
\affiliation{Max-Planck-Institut f\"ur Physik komplexer Systeme, 01187 Dresden, Germany}

\begin{abstract}
	Generic quantum many-body systems typically show a linear growth of the entanglement entropy after a quench from a product state. While entanglement is a property of the wave function, it is generated by the unitary time evolution operator and is therefore reflected in its increasing complexity as quantified by the \emph{operator entanglement entropy}.
Using numerical simulations of a static and a periodically driven quantum spin chain, we show that there is a robust correspondence between the entanglement entropy growth of \emph{typical} product states with the operator entanglement entropy of the unitary evolution operator, while \emph{special} product states, e.g. $\sigma_z$ basis states, can exhibit faster entanglement production.
In the presence of a disordered magnetic field in our spin chains, we show that both the wave function and operator entanglement entropies exhibit a power-law growth with the same disorder-dependent exponent, and clarify the apparent discrepancy in previous results. These systems, in the absence of conserved densities, provide further evidence for slow information spreading on the ergodic side of the many-body localization transition.
\end{abstract}

\maketitle

\section{Introduction}

Entanglement is an intrinsic property of quantum many-body wave functions and describes the amount of information a subsystem $A$ contains about its complement $B$. 
The entanglement between a bipartition $A::B$ of a closed quantum system is quantified by the entanglement entropy, which distinguishes separable product states with zero entanglement from states which are entangled to various degrees. For example, it was found that ground states of gapped quantum many-body systems exhibit an entanglement entropy which scales as the surface area of the subsystem $A$~\cite{Hastings:rev2007,Eisert:rev2010,Laflorencie:rev2016}. In generic quantum many-body systems that thermalize and in which the eigenstate thermalization hypothesis~\cite{Deutsch:1991,Srednicki:1994,Rigol:2008,Lazarides:2014,Alessio:2014,borgonovi_quantum_2016} is valid, the entanglement entropy of all eigenstates of the Hamiltonian is furthermore identified with the thermodynamic entropy of the subsystem and is therefore proportional to the \emph{volume} of the subsystem for eigenstates corresponding to finite temperatures~\cite{Santos:2011,Deutsch:2013,Beugeling:2015,Garrison:2018,Luitz_long:2016,Khaymovich:2019}.

The phenomenology becomes much richer in a nonequilibrium setting after a quantum quench: even if the wave function at early times is prepared as a lowly entangled state, the complex quantum many-body dynamics leads to a rapid production of entanglement and typically the entanglement entropy is found to grow linearly in time under the evolution with short range Hamiltonians~\cite{Calabrese:2005,Chiara:2006,Hyungwon:2013,Zhang:2015}. This was also observed in experiments with ultracold atomic systems~\cite{Kaufman_EEexp:2016}.

Although the above situation is true for a large class of generic quantum systems, strong disorder can completely destroy thermalization by inducing a phase transition to a many-body localized (MBL) phase \cite{Anderson:1958,Basko:2006,Gornyi:2005,Nandkishore:rev,Alet:rev,Abanin:rev,imbrie_diagonalization_2016,Alessio:2013,Lazarides:2015,Ponte:2015,Ponte:2015prl,Abanin:2016}. In the MBL phase -- which has also been realized in experiments~\cite{Schreiber:2015,Smith:2016} -- there is no transport of particles or energy density. However, surprisingly it was found that the entanglement entropy can grow logarithmically with time~\cite{Chiara:2006,Znidaric:2008,Bardarson:2012}, which was explained by an emergent set of quasi-local conserved quantities with residual interactions~\cite{serbyn_local_2013,serbyn_universal_2013,huse_phenomenology_2014}.

The production of entanglement is typically observed in the quench dynamics of wave functions $\ket{\psi(t)}$. However, it is uniquely due to the action of the time evolution operator $U(t)$ on the wave function and should therefore be reflected in the growing complexity of $U(t)$. Consequently, a generalization of the entanglement entropy to operators was introduced in \cite{Zinardi:2001,Prosen:2007}, which allows for a state-independent quantification of the complexity of operators across a bipartition of the system. In Refs.~\cite{Zhou:2017,Dubail:2017,Pal:2018}, this concept was directly applied to the time evolution operator $U(t)$, and a generic linear growth of the operator entanglement entropy (opEE) in ergodic quantum systems was found, while MBL systems are characterized by a logarithmic growth in time~\cite{Zhou:2017}.

In the vicinity of the MBL transition, the situation is more complicated: At intermediate disorder, where the system still thermalizes, an anomalously slow thermalization \cite{Luitz_long:2016,Luitz_anomalous:2016,Roy:2018,Herviou:2019,Baecker:2019,Colmenarez:2019arXiv} is found, which was connected 
to slow, subdiffusive transport~\cite{BarLev:2014,BarLev:2015,Agarwal:2015,Potter:2015,Vosk:2015,Znidaric:2016,Rehn:2016,BarLev:2017,Luitz_absence:2017,Bera:2017,Luitz_rev:2017,Agarwal_rev:2017,Kozarzewski:2018,Schulz:2018,Doggen:2018}. The anomalous thermalization is also reflected in a sublinear power-law growth~$\propto t^\alpha$ of the wave function entanglement entropy~\cite{Luitz_extended:2016}, even in systems which do not have globally conserved densities~\cite{Lezama:2019}, suggesting that the generic slow dynamics is a universal precursor of MBL. 
In such pre-MBL systems, the entanglement production exponent~$\alpha$ varies continuously with disorder and vanishes at the MBL transition, where the logarithmic growth takes over. The same phenomenology was found for the growth of the operator entanglement entropy of the time evolution operator $U(t)$, with a disorder-dependent exponent. Comparing the exponents found in the disordered Heisenberg model~\cite{Znidaric:2008,Berkelbach:2010,Pal:2010,Luitz:2015,BarLev:2015,Bera:2015}, of the entanglement entropy growth after a quench from a product state \cite{Luitz_extended:2016} and of the operator entanglement entropy growth of~$U(t)$~\cite{Zhou:2017}, it turns out that these exponents do not agree. 

In the present article, we address the above disagreement and show that it is due to the fact that completely unentangled $\sigma_z$ product states are not typical separable states, and that the growth of the operator entanglement entropy agrees perfectly with the growth of the wave function entanglement entropy if typical product states are considered.
We provide further evidence that the phenomenon of generally slow dynamics is universally found in one-dimensional systems at disorder strengths weak enough so that the system still thermalizes and does not require any conserved densities with associated transport.

Our article is organized as follows: In Sec.~\ref{sec:modmeth} we introduce a static and a periodically driven model representing generic disordered quantum XYZ spin chains, as well as the methods used to characterize them. 
We show that both models exhibit an MBL transition. The static model possesses no symmetries nor extensive conservation laws in addition to energy, while also the energy conservation is broken in the driven model. In Sec.~\ref{sec:framework} we present the framework of the entanglement measures considered for both wave functions and operators. In Sec.~\ref{sec:results} we compare the growth of the wave function entanglement entropy after a global quench,~$S_{\psi}(t)$, with the growth of the operator entanglement entropy of evolution operators, $S_{U}(t)$. For both models, we find the same universal power-law growth~$t^{\alpha}$ of~$S_{\psi}(t)$ and $S_{U}(t)$ within the ergodic regime--provided that $S_{\psi}(t)$ is obtained after a quench from typical initial product states. If the quench comes from either $\sigma_z$ product states or a class of intermediate states, $S_{\psi}(t)$ and $S_{U}(t)$ differ. Such a class of intermediate states is characterized by their maximal bond dimension in a matrix product state representation and a precise definition is given in Sec.~\ref{sec:initstate}. We further elucidate the influence of the initial states on the wave function entanglement production and argue that the underlying mechanism is that of monogamy of entanglement. 
Finally, in Sec.~\ref{sec:discussion} we conclude by summarizing and discussing our main results. 


\section{Model and Method} 
\label{sec:modmeth}

\subsection{Static and Driven XYZ chain}
We study the generic $\mathrm{XYZ}$ chain with open boundary conditions in the presence of a disordered tilted field. Its Hamiltonian is given by
\begin{align}
H_{0} = \sum_{i=1}^{L-1} & \left(J_{x}\sigma_{i}^{x}\sigma_{i+1}^{x}+J_{y}\sigma_{i}^{y}\sigma_{i+1}^{y}+J_{z}\sigma_{i}^{z}\sigma_{i+1}^{z}\right) \nonumber
\\ & + \sum_{i}^{L} h_{i}\left(h_{x} \sigma_{i}^{x}+h_{y} \sigma_{i}^{y}+h_{z} \sigma_{i}^{z}\right),
\label{hams}
\end{align}
where $\sigma^{\gamma}_{i}$ denote Pauli matrices, $J_{\gamma}$ coupling constants and $h_{\gamma} \equiv \tilde{h}_{\gamma}/|\vec{h}|$ fixed amplitudes of the tilted field $\vec{h}$; $\gamma \equiv x,y,z$. The field is fixed in the direction of $\vec{h}$ but its magnitude is disordered and taken from a box distribution $h_{i} \in [-W,W]$. We set $(J_{x,} J_{y},J_{z}) = (0.5,0.7,1.0)$ and  $(\tilde{h}_{x},\tilde{h}_{y},\tilde{h}_{z})= (0.95,1.0,1.1)$ to remove all the extensive conservation laws except energy conservation. This means that we can not further reduce the many-body Hilbert space and there is only one block of size $\text{dim}(\mathcal{H})=2^L$.

To also break the energy conservation, we subject the system~\eqref{hams} to monochromatic driving. The resulting Floquet model is defined by
\begin{align} 
H(t)= & H_{0} + H_{D}(t);  \nonumber  \\
H_{D}(t) = & \frac{A}{2}\sin(\omega t)  \sum_{i=1}^{L-1} \sigma_{i}^{z}\sigma_{i+1}^{z},
\label{hamf}
\end{align}
with driving period~$T\equiv 2\pi/\omega = 2$ and driving amplitude~$A=0.5$. The Floquet operator over one driving period is defined as
\begin{equation}
\hat{U}_{F}(T) \equiv \mathcal{T} e^{-i \int_{0}^{T} dt H_{D}(t)},
\label{fop}
\end{equation}
where~$\mathcal{T}$ denotes the time-ordering operator. 
Since $\hat U_{F}(T)$ is a unitary operator, its eigenvalues $\omega_n$ lie on the complex unit circle.

\begin{figure}[tb]
	\centering
	\includegraphics[width=\columnwidth]{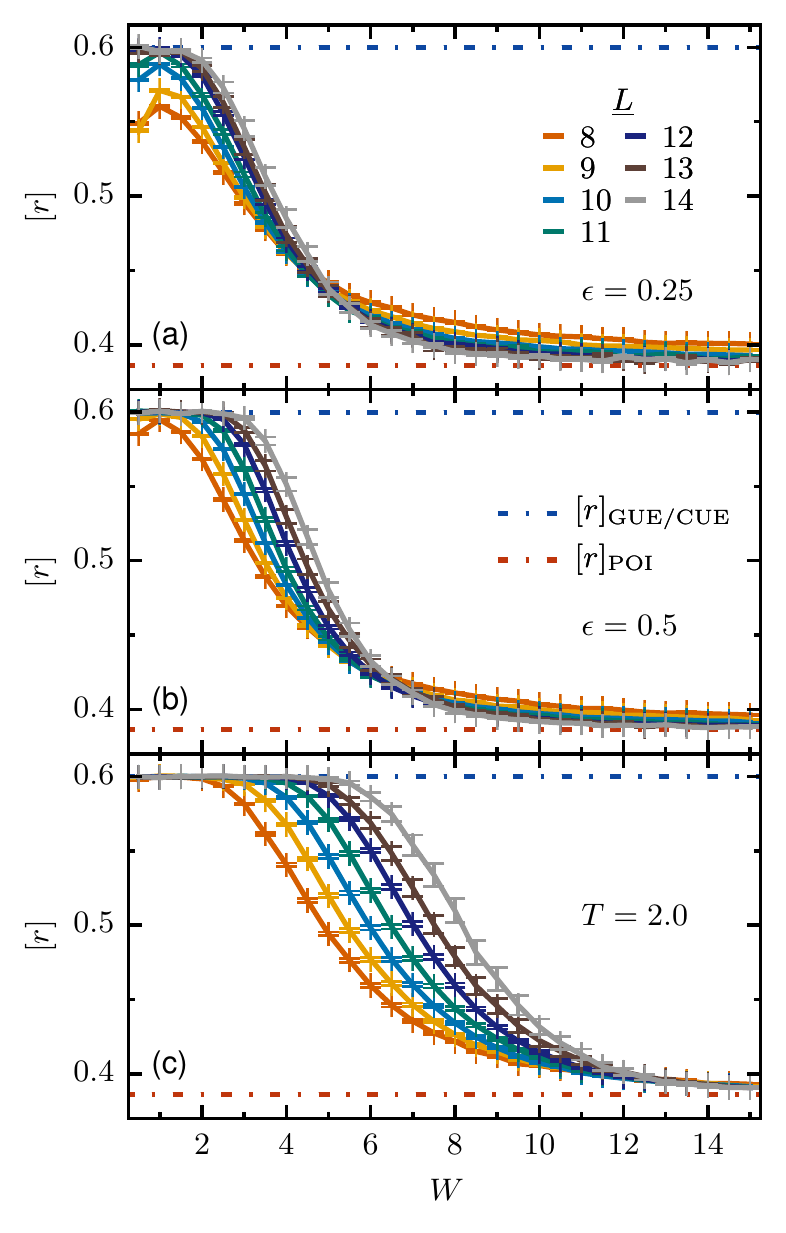}
	\caption{ Disorder-averaged level spacing ratio $[r]$ as a function of disorder strength $W$ and system size $L$. In the static model; for energy levels in a neighborhood of energy densities (a) $\epsilon = 0.25$ and (b) $\epsilon = 0.5$. (c) For all pairs of neighboring quasienergies in the Floquet model. 
		The legend of (a) applies to all panels.}
	\label{ls}
\end{figure}

\subsection{Method}

For the characterization of the models, we fully diagonalize the Hamiltonian of the static model~\eqref{hams} up to system size~$L=14$ with $\text{dim}(\mathcal H)=16384$, and consider the statistics of adjacent energy spacings. In the case of the Floquet model~\eqref{hamf}, we have to generate the time-evolution operator $\hat U_F(T)$. Since our driving protocol is monochromatic, we use small time steps $\mathrm{d}t=0.02$ and a second-order Trotter decomposition to separate the constant part $\exp(-\mathrm{i} \mathrm{d}t H_0)$ from the (diagonal) driven part $\exp(-\mathrm{i} \mathrm{d}t H_D(t))$, requiring only the diagonalization of $H_0$ to calculate the corresponding matrix exponential and repeated matrix products to step through the period $T$. We then fully diagonalize $\hat U_F(T)$ and consider the statistics of adjacent eigenphases to establish the ergodic regime of the model.

In the main part of this work, we consider both the production of wave function entanglement when starting from a product state $\ket{\psi(t=0)}$ as well as the growth of operator entanglement directly in the time evolution operator $U$. 

For the wave function dynamics, we use exact-time evolution with a Krylov-space method~\cite{nauts_new_1983,Saad:1992,moler_nineteen_2003,Luitz_extended:2016,Luitz_rev:2017}, which relies on the sparse structure of both~$H_0$ and~$H_D$. Again, to faithfully describe the monochromatic driving, small timesteps $\mathrm{d}t=0.02$ inside the period are used and the results analyzed including intraperiod values. This method allows for the calculation of entanglement entropies in systems up to $L=26$ with $\text{dim}(\mathcal H)=2^{26}\approx 6.7\cdot 10^7$.

For the operator entanglement entropy, we calculate the time evolution operator $U(t)$ as described above with the same limitation to system sizes up to $L=14$.

Our results are averaged over 50-100 disorder realizations, for several values of disorder strength within the ergodic regime at intermediate disorder $W\leq 4$.

\subsection{Characterization of the Model}
\label{sec:characterization}

The static model is similar to other disordered spin chains, in particular, the Heisenberg spin chain given by $J_x=J_y=J_z=1$, $h_x=h_y=0$ and $h_z=1$, which has in these units an MBL transition at $W_c\approx 7.5\pm 2.0$~\cite{Berkelbach:2010,Pal:2010,Luitz:2015,BarLev:2015,Bera:2015}, therefore we expect to find an MBL transition at similar values of disorder.

Using level spacing statistics as diagnosis, we corroborate that for the chosen parameters both models~\eqref{hams}~and~\eqref{hamf} undergo an MBL transition at critical values of disorder $W_{c}^{0} \approx 6$ and $W_{c}^{D} \approx 12$, respectively.

The adjacent level spacing ratio is defined as~\cite{Oganesyan:2007}
\begin{equation}
r = \frac{\mathrm{min}(\delta_{n}, \delta_{n+1})}{\mathrm{max}(\delta_{n},\delta_{n+1})},
\end{equation} 
where $\delta_{n}$ is defined in terms of consecutive energy levels $ \delta_{n} \equiv E_{n+1}-E_{n}$ of~\eqref{hams} or consecutive phase spacings $ \delta_{n} \equiv \theta_{n+1}-\theta_{n}$ extracted from the eigenvalues~$\omega_n = e^{-i\theta_{n}}$ of \eqref{fop}, depending on whether the static or the Floquet model is considered. 

For the static model~\eqref{hams} the energy is conserved and different energy densities $\epsilon$ represent different temperatures of the system. Since in the Heisenberg model numerical evidence for a mobility edge (i.e. an energy density-dependent critical disorder $W_c$) was found, we study the average level spacing ratio $[r]$ for different energy densities $\epsilon = (E_n-E_\text{min})/(E_\text{max}-E_\text{min})$ defined as in Ref. \cite{Luitz:2015}. We include the eigenvalues $E_n$ of the Hamiltonian on a vicinity of radius $\delta E=0.1$ around $\epsilon=0.5$ and $\epsilon=0.25$ and average over 100-10000 realizations of the disorder.
The results of this analysis for different system sizes $L$ are shown as a function of disorder strength~$W$ in~\figref{ls}(a),(b), revealing a transition from a mean level spacing expected from the Gaussian unitary ensemble (GUE) at weak disorder with $[r]_{\mathrm{GUE}} \approx 0.59982(8)$ \footnote{averaged over $10^5$ random GUE matrices of size $N=100$.} to the value predicted by a Poisson distribution $[r]_{\mathrm{POI}} \approx 0.38629$~\cite{Atas:2013} at strong disorder. Results for different system sizes show a crossing at intermediate disorder which, although the precise location still drifts with system size seems compatible with the existence of a mobility edge, as the crossing appears at significantly lower disorder for $\epsilon=0.25$ compared to~$\epsilon=0.5$.

In contrast, Floquet models which have an MBL transition do not exhibit mobility edges~\cite{Lazarides:2015,Roy:2018}. Therefore, we include eigenvalues from the full unit circle to calculate the level spacing statistics. The result is shown in~\figref{ls}(c), where  the crossing of $[r]$ with~$L$ is now enclosed between the limit values corresponding to the Poisson and the circular unitary ensemble (CUE) distributions,~$[r]_{\mathrm{POI}} \approx 0.38629$ and $[r]_{\mathrm{CUE}} \approx 0.59982(8)$~\footnote{We note that for large matrices, CUE and GUE ensembles are identical, which is not the case for smaller matrices \cite{Alessio:2014}.}, respectively.  As expected~\cite{Lazarides:2015}, the critical disorder of the Floquet model is larger than in the static case.
In the remainder of this paper, we will focus on the region of weak enough disorder such that both models are well located in the ergodic phase ($W\leq4$).

\begin{figure*}[tb!]
\flushleft
\includegraphics[width=\textwidth]{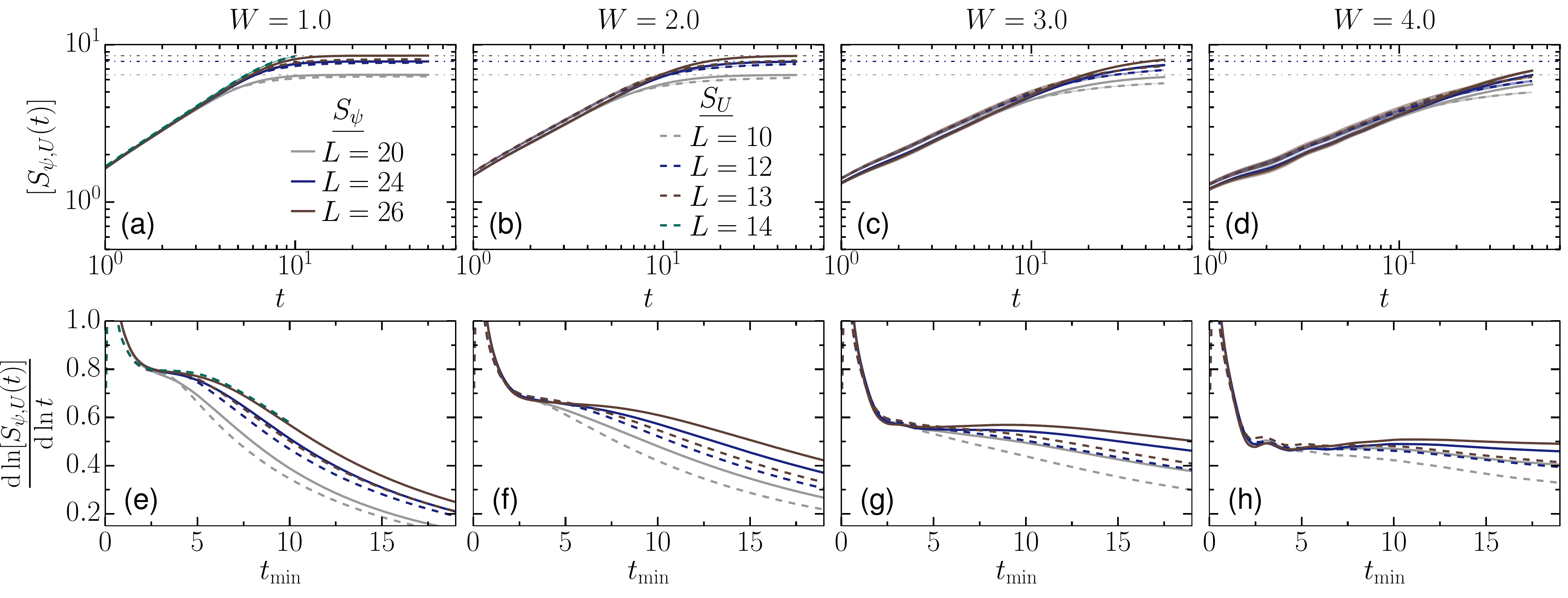}
\caption{Upper panel: Comparison between the disorder-averaged time evolution of the wave function entanglement entropy, $[S_{\psi}(t)]$ (solid lines), and the operator entanglement entropy, $[S_{U}(t)]$ (dashed lines); for the static model and for several values of disorder (a)-(d) on the ergodic side of the transition. Lower panel: The (discretized) logarithmic derivative of the data points in the upper panel, taken over time windows of size $\delta t =0.1$, starting from $t_{\mathrm{min}}$. The power-law behavior is visible by the emergent plateaus whose range increases with system size $L$. The legends in (a) and (b) apply to all panels.}
\label{fig:EE_opEE_static}
\end{figure*}

\section{Wave function and operator entanglement entropy}
 \label{sec:framework}
In this section, we provide a technical discussion pertinent to the concepts of the entanglement entropy of both quantum wave functions $\ket{\psi}$ and quantum evolution operators $\hat U$ for the case of a complementary real-space bipartition in terms of two subsystems $A$ and $B\equiv \overline A$, with a tensor product Hilbert space $\mathcal{H} = \mathcal{H}_{A} \otimes \mathcal{H}_{B}$. 
For a more detailed discussion we refer the reader to~\cite{Zinardi:2001,Zhou:2017,Dubail:2017}.

The system in a pure state $\ket{\psi}\in \mathcal{H}$ has a density matrix~$\rho = \ket{\psi} \bra{\psi}$ with unit purity $\tr \left(\rho^2 \right)=1$. The state of the subsystem~$A$ is described by the reduced density matrix $\rho_A$, given by the partial trace of~$\rho$ over the subsystem~$B$,
\begin{equation}
\rho_{A} =  \mathrm{Tr}_{B}(\rho).
\label{rdm}
\end{equation}
Even if $\rho$ is a pure state, the reduced density matrix is in general a mixed state with purity $\tr\left( \rho_A^2\right) \leq 1$.

The von Neumann entropy of the reduced density matrix $\rho_A$ is called the entanglement entropy and defined as
\begin{equation}
S_{\psi} = - \mathrm{Tr}\left(\rho_{A} \mathrm{ln} \rho_{A}\right).
\label{vne}
\end{equation}
Any pure state $\ket{\psi} \in \mathcal{H}$ in turn can be expressed in terms of its Schmidt decomposition
\begin{equation}
\ket{\psi} = \sum_{i}\sqrt{\lambda_{i}} \,\, \ket{\psi^{A}_{i}}\otimes \ket{\psi^{B}_{i}},
\label{ssd}
\end{equation}
where $\sqrt{\lambda_{i}}$ are the singular values of the matrix $\Psi\in\mathbb{C}^{\text{dim}(\mathcal{ H}_A) \times \text{dim}(\mathcal{H}_B)}$ with $\Psi_{i_A,i_B} = \braket{i_A,i_B|\psi}$, where $\ket{i_A,i_B}$ are computational basis states of $\mathcal H_A \otimes \mathcal H_B$.
The states $\{\ket{\psi^{A}_{i}} \}$ ($\{\ket{\psi^{B}_{i}} \}$) form a complete orthonormal basis of $\mathcal{H}_{A}$ ($\mathcal{H}_{B}$) and are obtained from the left and right singular vectors of the matrix $\Psi$. Algorithmically, (when using a complete computational basis ordered such that the tensor product structure is preserved), the entanglement spectrum $\set{\lambda_i}$ of a wave function $\ket{\psi}$ is obtained by reshaping the wave function into a matrix $\Psi$ such that the row indices correspond to basis states of subsystem $A$ and the column indices correspond to basis states of subsystem $B$. Then, a singular value decomposition (SVD) of $\Psi$ yields the singular values $\set{\sqrt{\lambda_i}}$.

The von Neumann entanglement entropy \eqref{vne} is readily generalized to the $\alpha$ R\'enyi entropy in terms of the squared singular values (the entanglement spectrum)~$\{\lambda_{i}\}$;
\begin{equation}
S^{\alpha}_{\psi} = \frac{1}{1-\alpha} \,\, \mathrm{ln} \sum_{i} \lambda^{\alpha}_{i},
\label{re}
\end{equation}
from which \eqref{vne} is recovered in the limit $\alpha \to 1$.

\begin{figure*}[tb!]
\flushleft
\includegraphics[width=\textwidth]{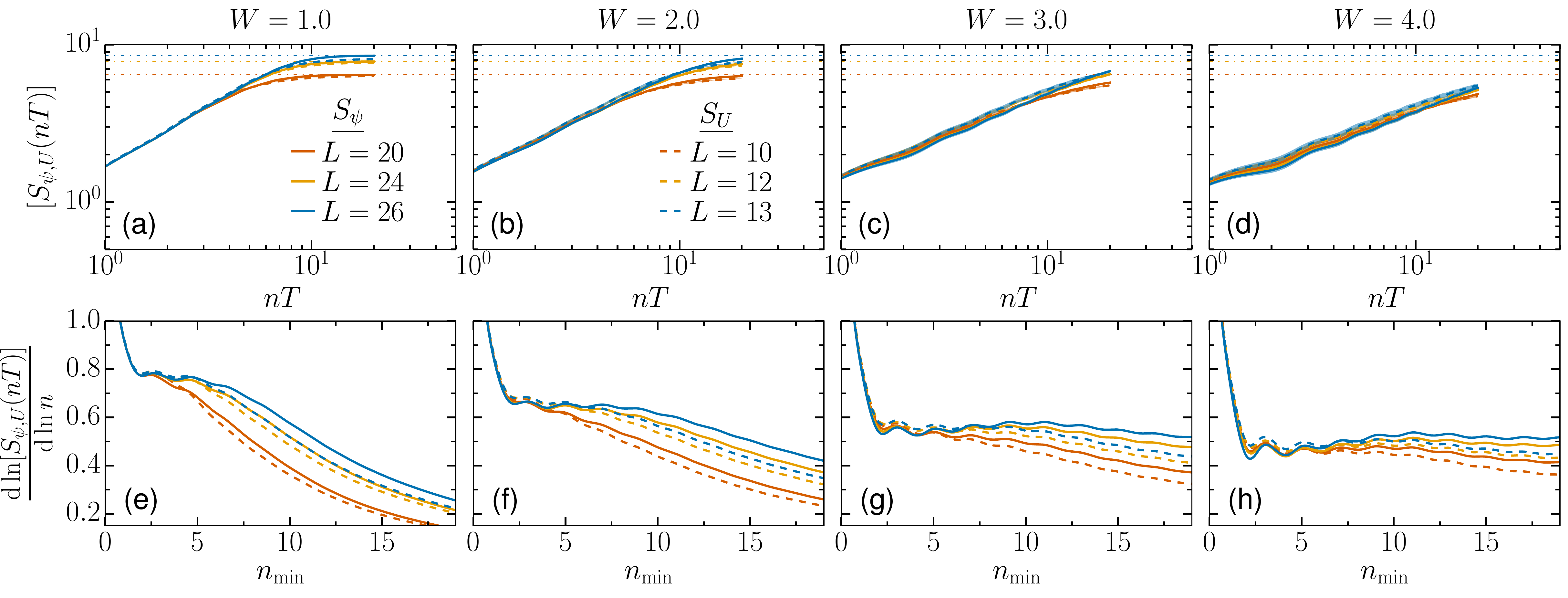}
\caption{Upper panel: Comparison between the disorder-averaged time evolution of the wave function entanglement entropy, $[S_{\psi}(nT)]$ (solid lines), and the operator entanglement entropy, $[S_{U}(nT)]$ (dashed lines); for the Floquet model and several values of disorder (a)-(d) on the ergodic side of the transition. Lower panel: The (discretized) logarithmic derivative of the data points in the upper panel, taken over intraperiod time windows of size $\delta t=0.1$, starting from $n_{\mathrm{min}}$. 
 The power-law behavior is visible by the plateaus whose range increases with system size $L$. The legends in (a) and (b) apply to all panels.}
\label{fig:EE_opEE_floquet}
\end{figure*}

From the foregoing concepts, the generalization of the entanglement entropy to the space of linear operators is straightforward. Linear operators $\hat{O}: \mathcal{H}\rightarrow\mathcal{H}$ form a basis of the Hilbert space $\tilde{\mathcal{H}}: \mathcal{H} \rightarrow \mathcal{H}$, endowed with the inner product $\langle \cdot,\cdot \rangle : \tilde{\mathcal{H}} \times \tilde{\mathcal{H}} \rightarrow \mathbb{C}$, which in turn is inherited from $\mathcal{H}$. Given two linear operators $\hat{O},\hat{O}^{'} \in \tilde{\mathcal{H}}$, their inner product is defined as 
\begin{equation}
\langle \hat{O},\hat{O}^{'} \rangle : = \frac{1}{\sqrt{\mathrm{dim(\tilde{\mathcal{H}})}}}\mathrm{Tr}\left(\hat{O}^{\dagger}\hat{O}^{'} \right).
\label{ip}
\end{equation}
The same extension can be done for the aforementioned bipartition ($A::B$) in terms of the Hilbert spaces associated to each subsystem, $\tilde{\mathcal{H}}_{A}: \mathcal{H}_{A} \rightarrow \mathcal{H}_{A}$ and $\tilde{\mathcal{H}}_{B}: \mathcal{H}_{B} \rightarrow \mathcal{H}_{B}$, respectively. Given any linear operator--in particular--a unitary quantum evolution operator
\begin{equation}
\hat{U} \equiv \mathcal{T}e^{-i\int_{0}^{t}dsH(s)},
\label{uo}
\end{equation} 
which obeys the orthonormality (unitarity) condition~$\langle\hat{U},\hat{U}\rangle = 1$, the above definitions \eqref{rdm}-\eqref{re} can be extended using~Eq.\eqref{ip}. 

Again, using a computational basis order which respects the tensor-product structure of the Hilbert space, we can write any basis state $\ket{i}=\ket{i_A,i_B}=\ket{i_A}\otimes \ket{i_B}$ in terms of basis states of the subsystems~$A$ and~$B$. Then, the time evolution operator~$\hat U$ has a matrix representation in the form $U_{i,j} = U_{(i_A,i_B),(j_A,j_B)} = \braket{i_A,i_B|\hat U |j_A,j_B}$. 

Similarly to the case of wave functions, but now dealing with two pairs of indices, we can calculate the operator entanglement spectrum by first vectorizing the matrix $U$ (interpreting it as a vector in $\tilde{ \mathcal H}\otimes\tilde{ \mathcal H}$) and then writing it as a matrix $u$ with all the indices corresponding to the $A$ subsystem as row indices and the indices corresponding to the $B$ subsystem as column indices:

\begin{equation}
U_{(i_A,i_B),(j_A,j_B)} \to u_{(i_A,j_A),(i_B,j_B)} \in \mathbb{C}^{\text{dim}(\mathcal H_A)^2 \times\text{dim}(\mathcal H_B)^2}.
\end{equation}

Algorithmically, this corresponds to interpreting the unitary matrix $U$ as a tensor of rank 4, then performing a tensor transposition to sort the $A$ and $B$ indices, followed by a reinterpretation as a (rectangular) matrix~$u$.
The \emph{operator entanglement spectrum} $\set{\lambda_i^{\text{op}}}$ is readily obtained by an SVD of $u$.

In the following, we will concern ourselves with comparing the entanglement dynamics of wave functions and time evolution operators $\hat U$ in the static \eqref{hams} and the Floquet model~\eqref{hamf}, taking into account the following consideration. Since there is a Hilbert space isomorphism between the space of states and the space of linear operators $ {\mathcal{H}}\otimes \mathcal{H} \cong \tilde{\mathcal{H}}$, the comparison between $S_{\psi}(t)$ and $S_{U}(t)$ should be done with respect to system sizes $L$ and~$L/2$, respectively, which correspond to the same Hilbert space dimension. The Hilbert space dimension determines the maximal entanglement entropy in a finite system, which is given by $S^\psi_\text{max} = \ln \text{min}\left[\text{dim}(\mathcal H_A),\text{dim}(\mathcal H_B)\right]$ for wave functions and $S^U_\text{max} = \ln \text{min}\left[\text{dim}(\tilde{\mathcal H}_A),\text{dim}(\tilde{\mathcal  H}_B)\right]=2S^\psi_\text{max}$ for operators, where the dynamics in ergodic systems is expected to reach values close to the maximum at late times~\cite{Page_average_1993}.

While in the case of the time evolution operator, there is no ambiguity with respect to the initial state ($\hat U(t=0)=\mathds{1}$ and therefore $S_U(t=0)=0$), we are free to choose any wave function $\ket{\psi(t=0)}$ as an initial state for the time evolution. Since we are interested in the production of entanglement, it is natural to require that the initial state be a product state $\ket{\psi(t=0)} = \ket{\psi_A}\otimes \ket{\psi_B}$ which has minimal entanglement $S_\psi(t=0)=0$. 

In this work, we mostly focus on typical product states, 
\begin{equation}
\ket{\psi_{AB}} \equiv \ket{\psi_{A}} \otimes \ket{\psi_{B}}
\label{abis}
\end{equation}
which are defined by random Haar measure states $\ket{\psi_A}$ and $\ket{\psi_B}$ on each subsystem. These states are the most general product states and have zero entanglement between $A$ and $B$, while being maximally entangled inside each subsystem. 
We will discuss the dependence on initial states for different classes of product states in Sec.~\ref{sec:initstate}, including the much simpler (and less typical) $\sigma_z$ basis states which are widely used in related numerical studies.

\section{Numerical results}
\label{sec:results}

\subsection{Static model}

In~\figref{fig:EE_opEE_static} we show the growth of the entanglement entropy of $\ket{\psi(t)}$ (solid lines) following a global quench from a typical product state of the form~\eqref{abis}, and compare it with the growth of the operator entanglement entropy of $U(t)$ (dashed lines) for disorder strengths $W=1,2,3,4$, all well in the ergodic regime, as discussed in Sec. \ref{sec:characterization}. All results are averaged over $\approx 100$ realizations of the disorder and for each disorder realization a different initial product state is used.

In the upper panels of Fig.~\ref{fig:EE_opEE_static} we observe that for all values of disorder (starting from zero entanglement) there is a rapid growth of both entropies at short times and a saturation at late times. The saturation values are close to the Page value (indicated by dotted-dashed lines), as expected~\cite{Zhou:2017}.
Interestingly, the growth of the operator entanglement entropy is almost identical to the wave function entanglement entropy growth, clearly showing that they encode the same information. 

The doubly logarithmic scale reveals that the growth of both entropies follows a power law $t^\alpha$ in time until saturation, and the domain of the power law extends to later times for larger systems due to the larger saturation value (proportional to $L$). 

To analyze the value of the dynamical exponent $\alpha$, we show the discretized logarithmic derivative $\mathrm{d} \ln S(t)/\mathrm{d} \ln t$ of both entropies as a function of time in the lower panels of Fig. \ref{fig:EE_opEE_static}, where the derivative is taken over small time windows of size $\delta t =0.1$ (results are essentially independent of the choice of $\delta t$).  This analysis clearly shows that both the operator entanglement entropy of~$U(t)$ and the wave function entanglement entropy of $\ket{\psi(t)}$ grow like a power law with the same exponent $\alpha$. The domain of the power law grows with system size, stabilizing the plateau in the logarithmic derivative. And, most importantly, the value of the exponent decreases continuously as a function of disorder, confirming that there is slow dynamics in the system before it undergoes an MBL transition. These results appear to be converged with system size at short enough times, and due to the slower dynamics the domain of the power law is larger for stronger disorder~$W$.

At weak disorder, the dynamical exponent $\alpha$ approaches the ballistic limit ($\alpha=1$) for clean nonintegrable systems~\cite{Chiara:2006,Hyungwon:2013,Zhang:2015} consistent with a saturated Lieb-Robison bound \cite{Lieb:1972}. 

\subsection{Floquet model}

Analogously to the case of the static model, in~\figref{fig:EE_opEE_floquet} we show the comparison between the  growth of the wave function entanglement entropy of typical product states~$\ket{\psi(t)}$ and the operator entanglement entropy of~$U(t)$, generated by the monochromatic drive~\eqref{hamf}. 

The results are very similar to the static case: We find a power law growth of both entropies at short times until it saturates to a value close to the Page value at late times. The domain of the power law grows with system size $L$ and the operator entanglement entropy of~$U(t)$ follows very closely the wave function entanglement entropy (comparing the entanglement entropy of a system of size $L$ to the opEE of a system of size $L/2$). 

We find again that at stronger disorder (still well in the ergodic regime), the exponent $\alpha$ of this power law is strongly suppressed, indicating slow dynamics in this system prior to the MBL transition.

\begin{figure}[tb]
\centering
\includegraphics[width=\columnwidth]{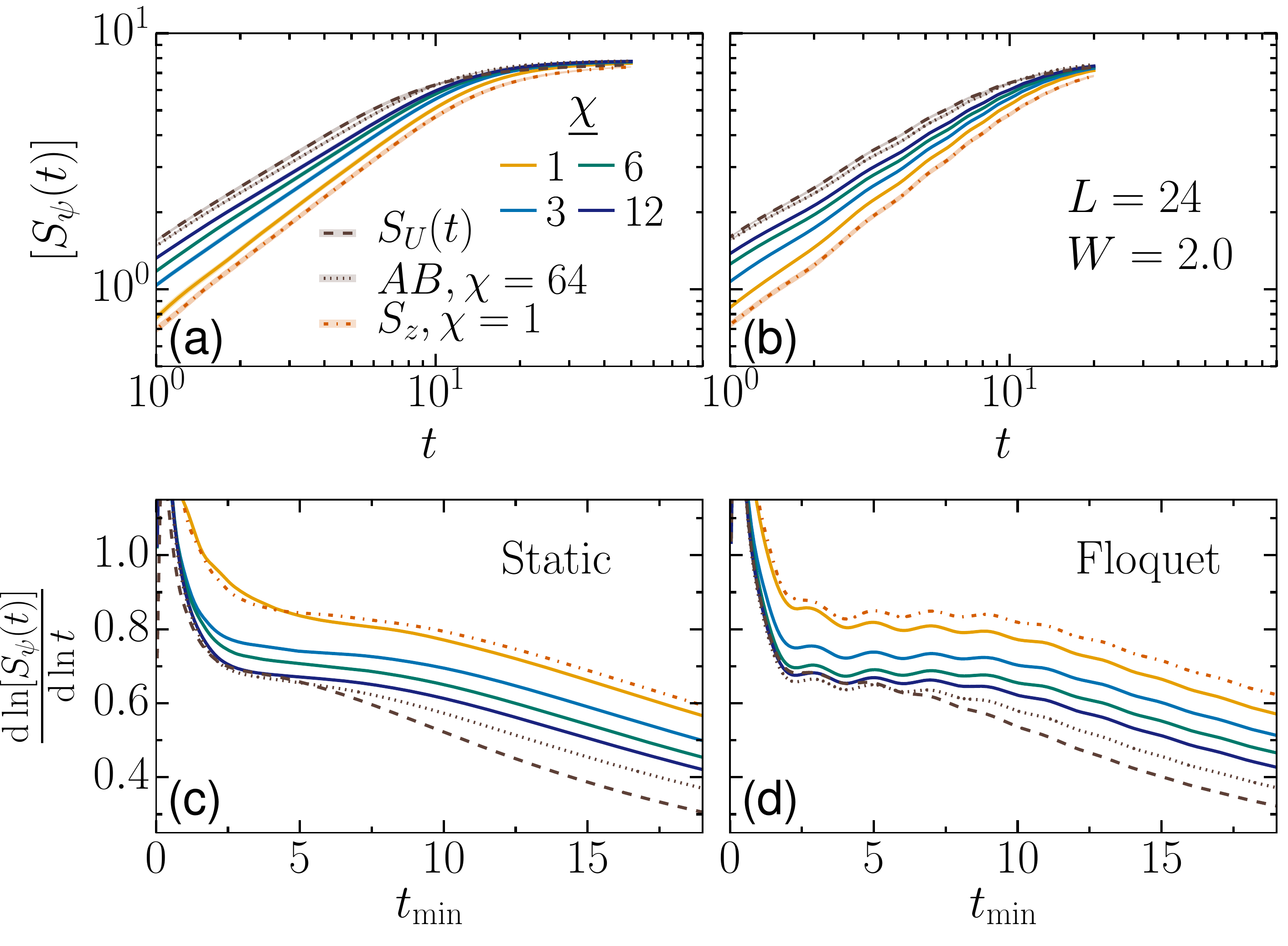}
\caption{Upper panels: Disorder-averaged entanglement entropy growth after a quench from $AB,\chi=64$, $\sigma_{z},\chi=1$, and intermediate random product initial states of bond dimension~$\chi$; in dashed lines the operator entanglement entropy growth, $S_{U}(t)$ for $L=12$. (a) For the static model, (b) for the Floquet model; $L=24$, and $W=2.0$. Lower panels: The (discretized) logarithmic derivative of the data points in the upper panel, taken over time windows of size $\delta t =0.1$, starting from $t_{\mathrm{min}}$, where (c) and (d) correspond to (a) and (b), respectively. The legends in (a) and (b) apply to all the panels.}
\label{fig:bdis}
\end{figure}

\begin{figure}[tb]
\centering
\includegraphics[width=\columnwidth]{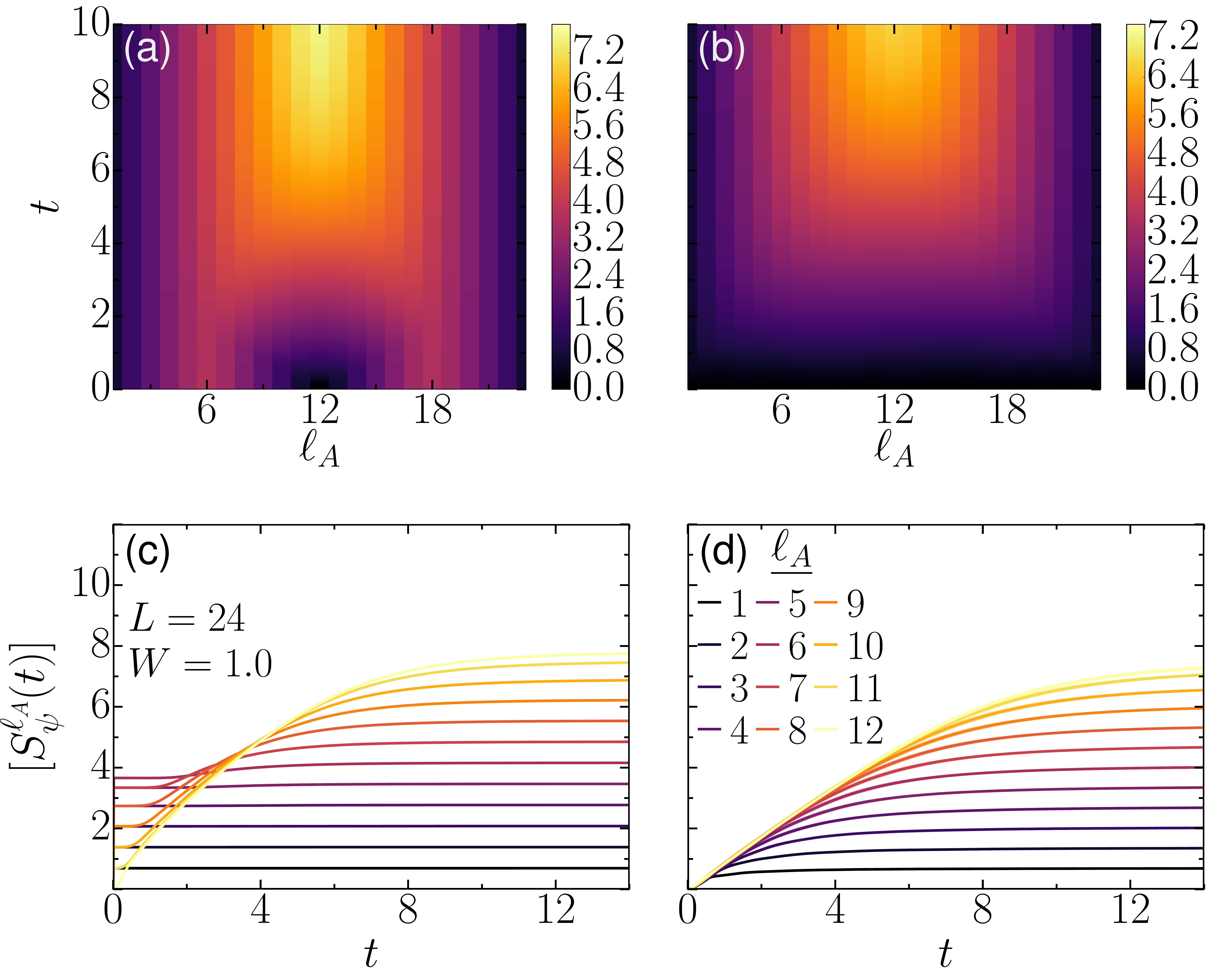}
\caption{Upper panels: Disorder-averaged entanglement entropy (in colorbar) as a function of the cut $\ell_{A} = 1-L$ and time; for the static model, $L=24$, and $W=1.0$. Following a quench from: (a) $AB,\chi=64$ product states and (b) $\sigma_{z},\chi=1$ product states. Lower panels: (c) sublinear and (d) linear entanglement entropy growth for the first half of the cuts $\ell_A=1-L/2$, corresponding to (a) and (b), respectively. The legends in (c) apply to all panels and legends in (d) apply to the lower panel.}
\label{fig:eecut}
\end{figure}

\subsection{Initial state dependence} 
\label{sec:initstate}

In Figs.~\ref{fig:EE_opEE_static} and \ref{fig:EE_opEE_floquet}, we have shown results for the entanglement entropy production starting from a typical product state, which is maximally entangled inside the subsystems $A$ and $B$. 

It is interesting to ask the question whether other classes of product states (which are less strongly entangled inside the subsystems) yield the same results. We therefore introduce the following general matrix product state (MPS) ansatz for any product state with respect to the $A :: B$ bipartition:

\begin{equation}
\begin{split}
 \ket{\psi_{\chi}} & = \frac{1}{\sqrt{\mathcal N}} \sum_{\set{\sigma}} \sum_{\{i\}=1}^\chi 
 M^{\sigma_{1}}_{i_{1}} M^{\sigma_{2}}_{i_{1}i_{2}} \cdots M^{\sigma_{\ell_A-1}}_{i_{\ell_A-1}i_{\ell_A}} M^{\sigma_{\ell_A}}_{i_{\ell_A}} \otimes  \\
  & M^{\sigma_{\ell_A+1}}_{i_{\ell_A+1}} M^{\sigma_{\ell_A+2}}_{i_{\ell_A+2}i_{\ell_A+3}}  \cdots M^{\sigma_{L-1}}_{i_{L-1}i_{L}} M^{\sigma_{L}}_{i_{L}}
|\sigma_{1}\cdots\sigma_{L}\rangle.
\end{split}
\label{eq:mps}
\end{equation}
Here, $\ell_A$ is the length of the subsystem $A$ and $M^{\sigma_k} \in \mathbb{C}^{\chi\times\chi}$ are independent random matrices with i.i.d Gaussian elements associated to each site and spin polarization of the system. At the edges $k=1, L$ of the system and at the boundary $k=\ell_A,\ell_A+1$, $M^{\sigma}_k \in \mathbb{C}^{\chi}$ are vectors instead, making the wave function a product state if cut at the boundary between $A$ and $B$. The wave function~$\ket{\psi_\chi}$ is normalized by a proper choice of the constant~$\mathcal N$.

The typical product states used in Figs.~\ref{fig:EE_opEE_static} and \ref{fig:EE_opEE_floquet} correspond to the maximal bond dimension $\chi=2^{\ell_A/2}$ (for even $\ell_A=L-\ell_A$).

If we choose instead $\chi=1$, we recover product states with random phases on each site but zero entanglement, independently of the bipartition of the system. The typical product states were also considered in~\cite{Jonay:2018}, and the special case of $\chi=1$ with and without random phases in Refs. \cite{Hyungwon:2013,Zhang:2015,Luitz_extended:2016,Lezama:2019}.

The general ansatz in Eq.~\eqref{eq:mps} allows for a smooth interpolation between these extreme cases by choosing different intermediate bond dimensions~$\chi$.

In Fig. \ref{fig:bdis}, we present the growth of the entanglement entropy in both the static and Floquet models for W=2.0 for different classes of initial product states defined by Eq.~\ref{eq:mps}, labelled by their bond dimensions $\chi$. In addition to $\chi=1$, we show data for the $\chi=1$ case without random phases, i.e. for pure $\sigma_z$ basis states.

Somewhat surprisingly, different classes of product states show a remarkably different behavior in terms of their entanglement growth even though we consider only the ergodic phase of our models. We note in passing that different (logarithmic) entanglement production rates were found in the MBL phase when considering different types of $\chi=1$ product states~\cite{Nanduri:2014}. 

The entanglement in completely unentangled product states grows the fastest, while it grows more slowly in typical product states which are maximally entangled within the subsystems. This different growth rate seems to be reflected in different exponents $\alpha$ of the power law in time.

We argue here that typical product states with maximal $\chi$ are the most representative for the overall behavior of the system in the sense that they contain the largest number of degrees of freedom. This means that if one generates a random product state, the likelihood of finding a $\chi=1$ state vanishes compared to a maximal $\chi$ state.

Why is the growth of entanglement slower if we start from a product state which is initially already entangled inside each subsystem, compared to an unentangled $\chi=1$ product state? The different behavior in the two cases is illustrated in Fig. \ref{fig:eecut}, where the top panels exhibit the entanglement entropy growth as a function of time for different sizes of the subsystem. Panel (a) shows the case of a typical product state, which has zero entanglement for $\ell_A=L/2$ by construction and maximal entanglement for all other values of $\ell_A$ given this constraint. Panel~(c) shows the same data as line plots for different $\ell_A$. It is clear that the entanglement entropy for a bipartition which is already close to maximally entangled can only grow very little. However, due to the monogamy of entanglement~\cite{Coffman:2000,Osborne:2006}, in order to generate entanglement across the cut at $\ell_A=L/2$, highly nontrivial processes have to occur to ``free'' some degrees of freedom before they can entangle with the other subsystem. This is reflected in the flat behavior (no growth) of cuts at e.g. $\ell_A=5$ at short times and slows down the entanglement growth across the cut at the centre.

Conversely, the case of a $\chi=1$ $\sigma_z$ product state does not impose such constraints and the entanglement for all cuts builds up homogeneously.

The entanglement production at very early times $t\ll1$ is different from the situation described above, the typical product states produce entanglement much faster, compared to the $\chi=1$ product states, and it is after this fast start that the entanglement production becomes slower. Interestingly, this $t\ll1$ behavior was observed for similar typical product states~\cite{Jonay:2018}, where the faster entanglement production was attributed to the positive curvature
$\partial^{2} S(\ell_{A},t)/ \partial \ell_{A}^{2}$ at the central cut $\ell_{A} = L/2$, compared to the flat structure produced by the $\chi=1$ initial condition. This situation is reflected in~\figref{fig:eecut} (a),(b) at $t\ll1$ around the central cut. The aforementioned second partial derivative is the subleading correction to the coarse-grained behavior of the local entanglement entropy increase rate and it is explained by the phenomenology of entanglement production~\cite{Nahum:2017,Nahum:2018,Jonay:2018}, which is conjectured to apply to generic non-integrable systems.  

\section{Discussion}
\label{sec:discussion}

We have systematically compared the operator entanglement entropy of the time evolution operator $U$ of disordered static and driven quantum spin chains, well in the ergodic regime of the phase diagram. Our models are chosen such that they exhibit a many-body localization transition at strong disorder.

It is known that such systems in general exhibit slow dynamics, most notably reflected in subdiffusive transport~\cite{BarLev:2014,BarLev:2015,Agarwal:2015,Potter:2015,Vosk:2015,Znidaric:2016,Rehn:2016,BarLev:2017,Luitz_anomalous:2016,Luitz_rev:2017,Agarwal_rev:2017,Kozarzewski:2018,Schulz:2018,Doggen:2018}, and it was found that slow dynamics is also visible in the spreading of quantum information~\cite{Luitz_extended:2016,Luitz_information:2017}, even in the absence of conserved quantities which could be transported \cite{Lezama:2019}.

In the case of the disordered quantum Heisenberg chain, the operator entanglement entropy of the time evolution operator was calculated in Ref.~\cite{Zhou:2017}, and a slow, sublinear power-law growth was found. However, comparing these exponents to the exponents of the power-law growth of the wave function entanglement entropy after a quench from a $\chi=1$ $\sigma_z$ product state revealed a discrepancy of these exponents.

Here, we clarify this discrepancy: Our argument given above based on the concept of monogamy of entanglement explains why $\chi=1$ product states can exhibit genuinely faster entanglement production compared to typical product states. However, since the time evolution operator needs to encode the entanglement production for any initial state, its entanglement entropy can only be expected to grow with a rate similar to that seen in typcial wave functions, which are given by initial product states with maximal bond dimension $\chi$ and represent an overwhelming majority in the class of all $A::B$ separable pure states.

This becomes even more apparent if we consider the time evolution operator $U$ in the computational $\sigma_z$ basis: Take an $\sigma_z$ product state $\ket{\phi}$, which is given by a single basis state $\ket{j}$: $\braket{i|\phi}=\phi_{i} = \delta_{i,j}$. Then, the time evolution of this state is given by $U_{ik} \phi_k  = U_{ij}$, which is the $j$-th column of $U$. On the other hand, when considering a product state with maximal $\chi$, we will get a random average over all columns of $U$, which will therefore reflect the typical behavior of all columns. 

At times $t\ll 1$ the opposite situation happens, the typical product states exhibit faster entanglement production compared to the $\chi=1$ product states.

In summary, we have clarified the apparent discrepancy between the growth of entanglement entropies of wave functions and time evolution operators in the slow dynamical regime prior to the MBL transition, showing that typical product states display identical power law entanglement entropy growth exponents compared with the operator entanglement entropy of the evolution operator. This is in agreement with the same correspondence recently observed in the case of linear growth in~\cite{Jonay:2018}. Furthermore, our results show that this correspondence is valid across the full ergodic phase and holds both for a static model with a mobility edge as well as for a periodically driven system.

Our results furthermore provide additional evidence that the slow dynamics in the ergodic phase is visible even in general situations of systems without conservation laws and can be observed in power-law growths of quantum information measures such as the operator entanglement entropy of the evolution operator and the entanglement production in wave functions.

\section{Acknowledgements}
We thank David A. Huse for useful comments on the manuscript and in particular for pointing us to the early time behavior.

\appendix

\section{R\'enyi operator entanglement entropies}

\begin{figure}[b!]
\flushleft
\includegraphics[width=\columnwidth]{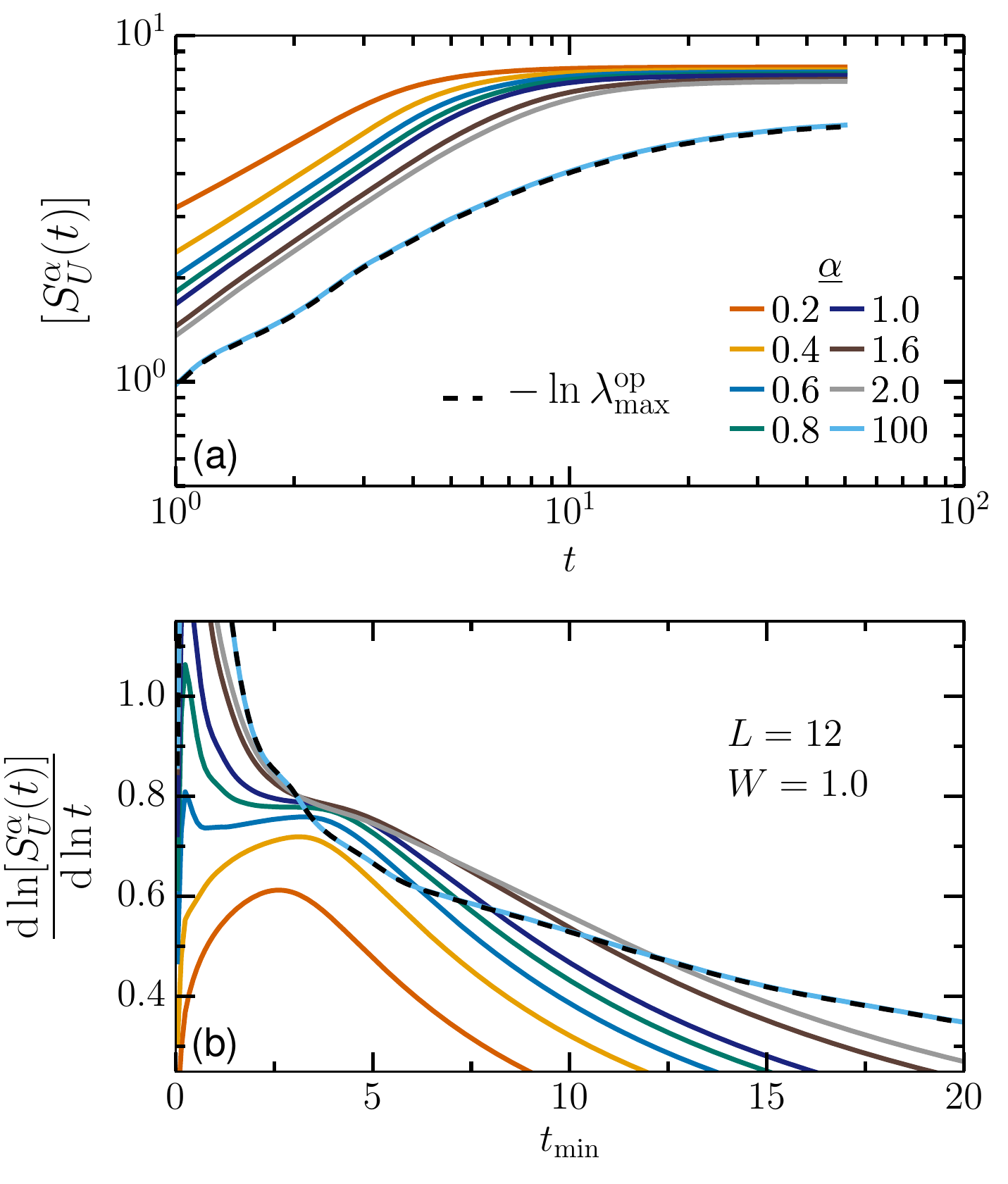}
\caption{(a) Disorder-averaged entanglement R\'enyi entropies $[S^{\alpha}_{U}(t)]$ (solid lines). In dashed lines $-\ln \lambda^{\mathrm{op}}_{\mathrm{max}}$ which is obtained in the limit $\alpha>>1$ ($\lambda^{\mathrm{op}}_{\mathrm{max}}$  denotes the maximum singular value of $S^{\alpha}_{U}(t)$); for the static model, $L=24$, and $W=2.0$. (b) The (discretized) logarithmic derivative of the data points in the upper panel, taken over time windows of size $\delta t =0.1$, starting from $t_{\mathrm{min}}$. The legends in (a) and (b) apply to both panels.}
\label{opEE_renyi_maxsv}
\end{figure}

In this appendix we provide further results on R\'enyi operator entanglement entropies. 
R\'enyi entropies are a generalization of the von Neumann-Shannon entropy. They highlight the behavior of different scales in the entanglement spectrum via the R\'enyi index $\alpha$. Here, we show results for the growth of the $\alpha$ R\'enyi operator entanglement entropy of the time evolution operator $U(t)$ of the static model in Eq. \eqref{hams}. In Fig. \ref{opEE_renyi_maxsv}, the top panel shows that the R\'enyi entropies grow for all values of $\alpha$. Large $\alpha$ highlights the behavior of the largest singular value, while smaller $\alpha>0$ represent their average behavior. The lower panel shows the discretized logarithmic derivative, which reveals that all R\'enyi entropies with $0.4\leq \alpha \leq 2$ essentially have the same power law growth exponent in time. For very large R\'enyi index $\alpha=100$, the behavior of the largest singular value is recovered and the entropy is identical to the limit $\alpha\to\infty$. Interestingly, in this case we observe a slightly slower entropy growth with a smaller exponent.


\bibliography{references}

\begin{thebibliography}{86}%
\makeatletter
\providecommand \@ifxundefined [1]{%
 \@ifx{#1\undefined}
}%
\providecommand \@ifnum [1]{%
 \ifnum #1\expandafter \@firstoftwo
 \else \expandafter \@secondoftwo
 \fi
}%
\providecommand \@ifx [1]{%
 \ifx #1\expandafter \@firstoftwo
 \else \expandafter \@secondoftwo
 \fi
}%
\providecommand \natexlab [1]{#1}%
\providecommand \enquote  [1]{``#1''}%
\providecommand \bibnamefont  [1]{#1}%
\providecommand \bibfnamefont [1]{#1}%
\providecommand \citenamefont [1]{#1}%
\providecommand \href@noop [0]{\@secondoftwo}%
\providecommand \href [0]{\begingroup \@sanitize@url \@href}%
\providecommand \@href[1]{\@@startlink{#1}\@@href}%
\providecommand \@@href[1]{\endgroup#1\@@endlink}%
\providecommand \@sanitize@url [0]{\catcode `\\12\catcode `\$12\catcode
  `\&12\catcode `\#12\catcode `\^12\catcode `\_12\catcode `\%12\relax}%
\providecommand \@@startlink[1]{}%
\providecommand \@@endlink[0]{}%
\providecommand \url  [0]{\begingroup\@sanitize@url \@url }%
\providecommand \@url [1]{\endgroup\@href {#1}{\urlprefix }}%
\providecommand \urlprefix  [0]{URL }%
\providecommand \Eprint [0]{\href }%
\providecommand \doibase [0]{http://dx.doi.org/}%
\providecommand \selectlanguage [0]{\@gobble}%
\providecommand \bibinfo  [0]{\@secondoftwo}%
\providecommand \bibfield  [0]{\@secondoftwo}%
\providecommand \translation [1]{[#1]}%
\providecommand \BibitemOpen [0]{}%
\providecommand \bibitemStop [0]{}%
\providecommand \bibitemNoStop [0]{.\EOS\space}%
\providecommand \EOS [0]{\spacefactor3000\relax}%
\providecommand \BibitemShut  [1]{\csname bibitem#1\endcsname}%
\let\auto@bib@innerbib\@empty
\bibitem [{\citenamefont {Hastings}(2007)}]{Hastings:rev2007}%
  \BibitemOpen
  \bibfield  {author} {\bibinfo {author} {\bibfnamefont {M~B}\ \bibnamefont
  {Hastings}},\ }\bibfield  {title} {\enquote {\bibinfo {title} {An area law
  for one-dimensional quantum systems},}\ }\href {\doibase
  10.1088/1742-5468/2007/08/p08024} {\bibfield  {journal} {\bibinfo  {journal}
  {Journal of Statistical Mechanics: Theory and Experiment}\ }\textbf {\bibinfo
  {volume} {2007}},\ \bibinfo {pages} {P08024--P08024} (\bibinfo {year}
  {2007})}\BibitemShut {NoStop}%
\bibitem [{\citenamefont {Eisert}\ \emph {et~al.}(2010)\citenamefont {Eisert},
  \citenamefont {Cramer},\ and\ \citenamefont {Plenio}}]{Eisert:rev2010}%
  \BibitemOpen
  \bibfield  {author} {\bibinfo {author} {\bibfnamefont {J.}~\bibnamefont
  {Eisert}}, \bibinfo {author} {\bibfnamefont {M.}~\bibnamefont {Cramer}}, \
  and\ \bibinfo {author} {\bibfnamefont {M.~B.}\ \bibnamefont {Plenio}},\
  }\bibfield  {title} {\enquote {\bibinfo {title} {Colloquium: Area laws for
  the entanglement entropy},}\ }\href {\doibase 10.1103/RevModPhys.82.277}
  {\bibfield  {journal} {\bibinfo  {journal} {Rev. Mod. Phys.}\ }\textbf
  {\bibinfo {volume} {82}},\ \bibinfo {pages} {277--306} (\bibinfo {year}
  {2010})}\BibitemShut {NoStop}%
\bibitem [{\citenamefont {Laflorencie}(2016)}]{Laflorencie:rev2016}%
  \BibitemOpen
  \bibfield  {author} {\bibinfo {author} {\bibfnamefont {Nicolas}\ \bibnamefont
  {Laflorencie}},\ }\bibfield  {title} {\enquote {\bibinfo {title} {Quantum
  entanglement in condensed matter systems},}\ }\href {\doibase
  https://doi.org/10.1016/j.physrep.2016.06.008} {\bibfield  {journal}
  {\bibinfo  {journal} {Physics Reports}\ }\textbf {\bibinfo {volume} {646}},\
  \bibinfo {pages} {1 -- 59} (\bibinfo {year} {2016})},\ \bibinfo {note}
  {quantum entanglement in condensed matter systems}\BibitemShut {NoStop}%
\bibitem [{\citenamefont {Deutsch}(1991)}]{Deutsch:1991}%
  \BibitemOpen
  \bibfield  {author} {\bibinfo {author} {\bibfnamefont {J~M}\ \bibnamefont
  {Deutsch}},\ }\bibfield  {title} {\enquote {\bibinfo {title} {{Quantum
  statistical mechanics in a closed system}},}\ }\href
  {http://link.aps.org/doi/10.1103/PhysRevA.43.2046} {\bibfield  {journal}
  {\bibinfo  {journal} {Phys. Rev. A}\ }\textbf {\bibinfo {volume} {43}},\
  \bibinfo {pages} {2046--2049} (\bibinfo {year} {1991})}\BibitemShut {NoStop}%
\bibitem [{\citenamefont {Srednicki}(1994)}]{Srednicki:1994}%
  \BibitemOpen
  \bibfield  {author} {\bibinfo {author} {\bibfnamefont {Mark}\ \bibnamefont
  {Srednicki}},\ }\bibfield  {title} {\enquote {\bibinfo {title} {{Chaos and
  quantum thermalization}},}\ }\href
  {http://link.aps.org/doi/10.1103/PhysRevE.50.888} {\bibfield  {journal}
  {\bibinfo  {journal} {Phys. Rev. E}\ }\textbf {\bibinfo {volume} {50}},\
  \bibinfo {pages} {888--901} (\bibinfo {year} {1994})}\BibitemShut {NoStop}%
\bibitem [{\citenamefont {Rigol}\ \emph {et~al.}(2008)\citenamefont {Rigol},
  \citenamefont {Dunjko},\ and\ \citenamefont {Olshanii}}]{Rigol:2008}%
  \BibitemOpen
  \bibfield  {author} {\bibinfo {author} {\bibfnamefont {Marcos}\ \bibnamefont
  {Rigol}}, \bibinfo {author} {\bibfnamefont {Vanja}\ \bibnamefont {Dunjko}}, \
  and\ \bibinfo {author} {\bibfnamefont {Maxim}\ \bibnamefont {Olshanii}},\
  }\bibfield  {title} {\enquote {\bibinfo {title} {{Thermalization and its
  mechanism for generic isolated quantum systems}},}\ }\href
  {http://www.nature.com/doifinder/10.1038/nature06838} {\bibfield  {journal}
  {\bibinfo  {journal} {Nature}\ }\textbf {\bibinfo {volume} {452}},\ \bibinfo
  {pages} {854--858} (\bibinfo {year} {2008})}\BibitemShut {NoStop}%
\bibitem [{\citenamefont {Lazarides}\ \emph {et~al.}(2014)\citenamefont
  {Lazarides}, \citenamefont {Das},\ and\ \citenamefont
  {Moessner}}]{Lazarides:2014}%
  \BibitemOpen
  \bibfield  {author} {\bibinfo {author} {\bibfnamefont {Achilleas}\
  \bibnamefont {Lazarides}}, \bibinfo {author} {\bibfnamefont {Arnab}\
  \bibnamefont {Das}}, \ and\ \bibinfo {author} {\bibfnamefont {Roderich}\
  \bibnamefont {Moessner}},\ }\bibfield  {title} {\enquote {\bibinfo {title}
  {Equilibrium states of generic quantum systems subject to periodic
  driving},}\ }\href {\doibase 10.1103/PhysRevE.90.012110} {\bibfield
  {journal} {\bibinfo  {journal} {Phys. Rev. E}\ }\textbf {\bibinfo {volume}
  {90}},\ \bibinfo {pages} {012110} (\bibinfo {year} {2014})}\BibitemShut
  {NoStop}%
\bibitem [{\citenamefont {D'Alessio}\ and\ \citenamefont
  {Rigol}(2014)}]{Alessio:2014}%
  \BibitemOpen
  \bibfield  {author} {\bibinfo {author} {\bibfnamefont {Luca}\ \bibnamefont
  {D'Alessio}}\ and\ \bibinfo {author} {\bibfnamefont {Marcos}\ \bibnamefont
  {Rigol}},\ }\bibfield  {title} {\enquote {\bibinfo {title} {Long-time
  behavior of isolated periodically driven interacting lattice systems},}\
  }\href {\doibase 10.1103/PhysRevX.4.041048} {\bibfield  {journal} {\bibinfo
  {journal} {Phys. Rev. X}\ }\textbf {\bibinfo {volume} {4}},\ \bibinfo {pages}
  {041048} (\bibinfo {year} {2014})}\BibitemShut {NoStop}%
\bibitem [{\citenamefont {Borgonovi}\ \emph {et~al.}(2016)\citenamefont
  {Borgonovi}, \citenamefont {Izrailev}, \citenamefont {Santos},\ and\
  \citenamefont {Zelevinsky}}]{borgonovi_quantum_2016}%
  \BibitemOpen
  \bibfield  {author} {\bibinfo {author} {\bibfnamefont {F.}~\bibnamefont
  {Borgonovi}}, \bibinfo {author} {\bibfnamefont {F.~M.}\ \bibnamefont
  {Izrailev}}, \bibinfo {author} {\bibfnamefont {L.~F.}\ \bibnamefont
  {Santos}}, \ and\ \bibinfo {author} {\bibfnamefont {V.~G.}\ \bibnamefont
  {Zelevinsky}},\ }\bibfield  {title} {\enquote {\bibinfo {title} {Quantum
  chaos and thermalization in isolated systems of interacting particles},}\
  }\href {\doibase 10.1016/j.physrep.2016.02.005} {\bibfield  {journal}
  {\bibinfo  {journal} {Physics Reports}\ }\textbf {\bibinfo {volume} {626}},\
  \bibinfo {pages} {1--58} (\bibinfo {year} {2016})}\BibitemShut {NoStop}%
\bibitem [{\citenamefont {Santos}\ \emph {et~al.}(2011)\citenamefont {Santos},
  \citenamefont {Polkovnikov},\ and\ \citenamefont {Rigol}}]{Santos:2011}%
  \BibitemOpen
  \bibfield  {author} {\bibinfo {author} {\bibfnamefont {Lea~F.}\ \bibnamefont
  {Santos}}, \bibinfo {author} {\bibfnamefont {Anatoli}\ \bibnamefont
  {Polkovnikov}}, \ and\ \bibinfo {author} {\bibfnamefont {Marcos}\
  \bibnamefont {Rigol}},\ }\bibfield  {title} {\enquote {\bibinfo {title}
  {Entropy of isolated quantum systems after a quench},}\ }\href {\doibase
  10.1103/PhysRevLett.107.040601} {\bibfield  {journal} {\bibinfo  {journal}
  {Phys. Rev. Lett.}\ }\textbf {\bibinfo {volume} {107}},\ \bibinfo {pages}
  {040601} (\bibinfo {year} {2011})}\BibitemShut {NoStop}%
\bibitem [{\citenamefont {Deutsch}\ \emph {et~al.}(2013)\citenamefont
  {Deutsch}, \citenamefont {Li},\ and\ \citenamefont {Sharma}}]{Deutsch:2013}%
  \BibitemOpen
  \bibfield  {author} {\bibinfo {author} {\bibfnamefont {J.~M.}\ \bibnamefont
  {Deutsch}}, \bibinfo {author} {\bibfnamefont {Haibin}\ \bibnamefont {Li}}, \
  and\ \bibinfo {author} {\bibfnamefont {Auditya}\ \bibnamefont {Sharma}},\
  }\bibfield  {title} {\enquote {\bibinfo {title} {Microscopic origin of
  thermodynamic entropy in isolated systems},}\ }\href {\doibase
  10.1103/PhysRevE.87.042135} {\bibfield  {journal} {\bibinfo  {journal} {Phys.
  Rev. E}\ }\textbf {\bibinfo {volume} {87}},\ \bibinfo {pages} {042135}
  (\bibinfo {year} {2013})}\BibitemShut {NoStop}%
\bibitem [{\citenamefont {Beugeling}\ \emph {et~al.}(2015)\citenamefont
  {Beugeling}, \citenamefont {Andreanov},\ and\ \citenamefont
  {Haque}}]{Beugeling:2015}%
  \BibitemOpen
  \bibfield  {author} {\bibinfo {author} {\bibfnamefont {W}~\bibnamefont
  {Beugeling}}, \bibinfo {author} {\bibfnamefont {A}~\bibnamefont {Andreanov}},
  \ and\ \bibinfo {author} {\bibfnamefont {Masudul}\ \bibnamefont {Haque}},\
  }\bibfield  {title} {\enquote {\bibinfo {title} {Global characteristics of
  all eigenstates of local many-body hamiltonians: participation ratio and
  entanglement entropy},}\ }\href {\doibase 10.1088/1742-5468/2015/02/p02002}
  {\bibfield  {journal} {\bibinfo  {journal} {Journal of Statistical Mechanics:
  Theory and Experiment}\ }\textbf {\bibinfo {volume} {2015}},\ \bibinfo
  {pages} {P02002} (\bibinfo {year} {2015})}\BibitemShut {NoStop}%
\bibitem [{\citenamefont {Garrison}\ and\ \citenamefont
  {Grover}(2018)}]{Garrison:2018}%
  \BibitemOpen
  \bibfield  {author} {\bibinfo {author} {\bibfnamefont {James~R.}\
  \bibnamefont {Garrison}}\ and\ \bibinfo {author} {\bibfnamefont {Tarun}\
  \bibnamefont {Grover}},\ }\bibfield  {title} {\enquote {\bibinfo {title}
  {Does a single eigenstate encode the full hamiltonian?}}\ }\href {\doibase
  10.1103/PhysRevX.8.021026} {\bibfield  {journal} {\bibinfo  {journal} {Phys.
  Rev. X}\ }\textbf {\bibinfo {volume} {8}},\ \bibinfo {pages} {021026}
  (\bibinfo {year} {2018})}\BibitemShut {NoStop}%
\bibitem [{\citenamefont {Luitz}(2016)}]{Luitz_long:2016}%
  \BibitemOpen
  \bibfield  {author} {\bibinfo {author} {\bibfnamefont {David~J.}\
  \bibnamefont {Luitz}},\ }\bibfield  {title} {\enquote {\bibinfo {title} {Long
  tail distributions near the many-body localization transition},}\ }\href
  {\doibase 10.1103/PhysRevB.93.134201} {\bibfield  {journal} {\bibinfo
  {journal} {Phys. Rev. B}\ }\textbf {\bibinfo {volume} {93}},\ \bibinfo
  {pages} {134201} (\bibinfo {year} {2016})}\BibitemShut {NoStop}%
\bibitem [{\citenamefont {Khaymovich}\ \emph {et~al.}(2019)\citenamefont
  {Khaymovich}, \citenamefont {Haque},\ and\ \citenamefont
  {McClarty}}]{Khaymovich:2019}%
  \BibitemOpen
  \bibfield  {author} {\bibinfo {author} {\bibfnamefont {Ivan~M.}\ \bibnamefont
  {Khaymovich}}, \bibinfo {author} {\bibfnamefont {Masudul}\ \bibnamefont
  {Haque}}, \ and\ \bibinfo {author} {\bibfnamefont {Paul~A.}\ \bibnamefont
  {McClarty}},\ }\bibfield  {title} {\enquote {\bibinfo {title} {Eigenstate
  thermalization, random matrix theory, and behemoths},}\ }\href {\doibase
  10.1103/PhysRevLett.122.070601} {\bibfield  {journal} {\bibinfo  {journal}
  {Phys. Rev. Lett.}\ }\textbf {\bibinfo {volume} {122}},\ \bibinfo {pages}
  {070601} (\bibinfo {year} {2019})}\BibitemShut {NoStop}%
\bibitem [{\citenamefont {Calabrese}\ and\ \citenamefont
  {Cardy}(2005)}]{Calabrese:2005}%
  \BibitemOpen
  \bibfield  {author} {\bibinfo {author} {\bibfnamefont {Pasquale}\
  \bibnamefont {Calabrese}}\ and\ \bibinfo {author} {\bibfnamefont {John}\
  \bibnamefont {Cardy}},\ }\bibfield  {title} {\enquote {\bibinfo {title}
  {Evolution of entanglement entropy in one-dimensional systems},}\ }\href
  {\doibase 10.1088/1742-5468/2005/04/p04010} {\bibfield  {journal} {\bibinfo
  {journal} {Journal of Statistical Mechanics: Theory and Experiment}\ }\textbf
  {\bibinfo {volume} {2005}},\ \bibinfo {pages} {P04010} (\bibinfo {year}
  {2005})}\BibitemShut {NoStop}%
\bibitem [{\citenamefont {Chiara}\ \emph {et~al.}(2006)\citenamefont {Chiara},
  \citenamefont {Montangero}, \citenamefont {Calabrese},\ and\ \citenamefont
  {Fazio}}]{Chiara:2006}%
  \BibitemOpen
  \bibfield  {author} {\bibinfo {author} {\bibfnamefont {Gabriele~De}\
  \bibnamefont {Chiara}}, \bibinfo {author} {\bibfnamefont {Simone}\
  \bibnamefont {Montangero}}, \bibinfo {author} {\bibfnamefont {Pasquale}\
  \bibnamefont {Calabrese}}, \ and\ \bibinfo {author} {\bibfnamefont {Rosario}\
  \bibnamefont {Fazio}},\ }\bibfield  {title} {\enquote {\bibinfo {title}
  {Entanglement entropy dynamics of heisenberg chains},}\ }\href {\doibase
  10.1088/1742-5468/2006/03/p03001} {\bibfield  {journal} {\bibinfo  {journal}
  {Journal of Statistical Mechanics: Theory and Experiment}\ }\textbf {\bibinfo
  {volume} {2006}},\ \bibinfo {pages} {P03001--P03001} (\bibinfo {year}
  {2006})}\BibitemShut {NoStop}%
\bibitem [{\citenamefont {Kim}\ and\ \citenamefont
  {Huse}(2013)}]{Hyungwon:2013}%
  \BibitemOpen
  \bibfield  {author} {\bibinfo {author} {\bibfnamefont {Hyungwon}\
  \bibnamefont {Kim}}\ and\ \bibinfo {author} {\bibfnamefont {David~A.}\
  \bibnamefont {Huse}},\ }\bibfield  {title} {\enquote {\bibinfo {title}
  {Ballistic spreading of entanglement in a diffusive nonintegrable system},}\
  }\href {\doibase 10.1103/PhysRevLett.111.127205} {\bibfield  {journal}
  {\bibinfo  {journal} {Phys. Rev. Lett.}\ }\textbf {\bibinfo {volume} {111}},\
  \bibinfo {pages} {127205} (\bibinfo {year} {2013})}\BibitemShut {NoStop}%
\bibitem [{\citenamefont {Zhang}\ \emph {et~al.}(2015)\citenamefont {Zhang},
  \citenamefont {Kim},\ and\ \citenamefont {Huse}}]{Zhang:2015}%
  \BibitemOpen
  \bibfield  {author} {\bibinfo {author} {\bibfnamefont {Liangsheng}\
  \bibnamefont {Zhang}}, \bibinfo {author} {\bibfnamefont {Hyungwon}\
  \bibnamefont {Kim}}, \ and\ \bibinfo {author} {\bibfnamefont {David~A.}\
  \bibnamefont {Huse}},\ }\bibfield  {title} {\enquote {\bibinfo {title}
  {Thermalization of entanglement},}\ }\href {\doibase
  10.1103/PhysRevE.91.062128} {\bibfield  {journal} {\bibinfo  {journal} {Phys.
  Rev. E}\ }\textbf {\bibinfo {volume} {91}},\ \bibinfo {pages} {062128}
  (\bibinfo {year} {2015})}\BibitemShut {NoStop}%
\bibitem [{\citenamefont {Kaufman}\ \emph {et~al.}(2016)\citenamefont
  {Kaufman}, \citenamefont {Tai}, \citenamefont {Lukin}, \citenamefont
  {Rispoli}, \citenamefont {Schittko}, \citenamefont {Preiss},\ and\
  \citenamefont {Greiner}}]{Kaufman_EEexp:2016}%
  \BibitemOpen
  \bibfield  {author} {\bibinfo {author} {\bibfnamefont {Adam~M.}\ \bibnamefont
  {Kaufman}}, \bibinfo {author} {\bibfnamefont {M.~Eric}\ \bibnamefont {Tai}},
  \bibinfo {author} {\bibfnamefont {Alexander}\ \bibnamefont {Lukin}}, \bibinfo
  {author} {\bibfnamefont {Matthew}\ \bibnamefont {Rispoli}}, \bibinfo {author}
  {\bibfnamefont {Robert}\ \bibnamefont {Schittko}}, \bibinfo {author}
  {\bibfnamefont {Philipp~M.}\ \bibnamefont {Preiss}}, \ and\ \bibinfo {author}
  {\bibfnamefont {Markus}\ \bibnamefont {Greiner}},\ }\bibfield  {title}
  {\enquote {\bibinfo {title} {Quantum thermalization through entanglement in
  an isolated many-body system},}\ }\href {\doibase 10.1126/science.aaf6725}
  {\bibfield  {journal} {\bibinfo  {journal} {Science}\ }\textbf {\bibinfo
  {volume} {353}},\ \bibinfo {pages} {794--800} (\bibinfo {year} {2016})},\
  \Eprint
  {http://arxiv.org/abs/https://science.sciencemag.org/content/353/6301/794}
  {https://science.sciencemag.org/content/353/6301/794} \BibitemShut {NoStop}%
\bibitem [{\citenamefont {Anderson}(1958)}]{Anderson:1958}%
  \BibitemOpen
  \bibfield  {author} {\bibinfo {author} {\bibfnamefont {P.~W.}\ \bibnamefont
  {Anderson}},\ }\bibfield  {title} {\enquote {\bibinfo {title} {Absence of
  diffusion in certain random lattices},}\ }\href {\doibase
  10.1103/PhysRev.109.1492} {\bibfield  {journal} {\bibinfo  {journal} {Phys.
  Rev.}\ }\textbf {\bibinfo {volume} {109}},\ \bibinfo {pages} {1492--1505}
  (\bibinfo {year} {1958})}\BibitemShut {NoStop}%
\bibitem [{\citenamefont {Basko}\ \emph {et~al.}(2006)\citenamefont {Basko},
  \citenamefont {Aleiner},\ and\ \citenamefont {Altshuler}}]{Basko:2006}%
  \BibitemOpen
  \bibfield  {author} {\bibinfo {author} {\bibfnamefont {D~M}\ \bibnamefont
  {Basko}}, \bibinfo {author} {\bibfnamefont {I~L}\ \bibnamefont {Aleiner}}, \
  and\ \bibinfo {author} {\bibfnamefont {B~L}\ \bibnamefont {Altshuler}},\
  }\bibfield  {title} {\enquote {\bibinfo {title} {{Metal{\textendash}insulator
  transition in a weakly interacting many-electron system with localized
  single-particle states}},}\ }\href
  {http://linkinghub.elsevier.com/retrieve/pii/S0003491605002630} {\bibfield
  {journal} {\bibinfo  {journal} {Ann. Phys.}\ }\textbf {\bibinfo {volume}
  {321}},\ \bibinfo {pages} {1126--1205} (\bibinfo {year} {2006})}\BibitemShut
  {NoStop}%
\bibitem [{\citenamefont {Gornyi}\ \emph {et~al.}(2005)\citenamefont {Gornyi},
  \citenamefont {Mirlin},\ and\ \citenamefont {Polyakov}}]{Gornyi:2005}%
  \BibitemOpen
  \bibfield  {author} {\bibinfo {author} {\bibfnamefont {I~V}\ \bibnamefont
  {Gornyi}}, \bibinfo {author} {\bibfnamefont {A~D}\ \bibnamefont {Mirlin}}, \
  and\ \bibinfo {author} {\bibfnamefont {D~G}\ \bibnamefont {Polyakov}},\
  }\bibfield  {title} {\enquote {\bibinfo {title} {{Interacting Electrons in
  Disordered Wires: Anderson Localization and Low-T Transport}},}\ }\href
  {http://journals.aps.org/prl/abstract/10.1103/PhysRevLett.95.206603}
  {\bibfield  {journal} {\bibinfo  {journal} {Phys. Rev. Lett.}\ }\textbf
  {\bibinfo {volume} {95}},\ \bibinfo {pages} {206603} (\bibinfo {year}
  {2005})}\BibitemShut {NoStop}%
\bibitem [{\citenamefont {Nandkishore}\ and\ \citenamefont
  {Huse}(2015)}]{Nandkishore:rev}%
  \BibitemOpen
  \bibfield  {author} {\bibinfo {author} {\bibfnamefont {Rahul}\ \bibnamefont
  {Nandkishore}}\ and\ \bibinfo {author} {\bibfnamefont {David~A.}\
  \bibnamefont {Huse}},\ }\bibfield  {title} {\enquote {\bibinfo {title}
  {Many-body localization and thermalization in quantum statistical
  mechanics},}\ }\href {\doibase 10.1146/annurev-conmatphys-031214-014726}
  {\bibfield  {journal} {\bibinfo  {journal} {Annual Review of Condensed Matter
  Physics}\ }\textbf {\bibinfo {volume} {6}},\ \bibinfo {pages} {15--38}
  (\bibinfo {year} {2015})},\ \Eprint
  {http://arxiv.org/abs/https://doi.org/10.1146/annurev-conmatphys-031214-014726}
  {https://doi.org/10.1146/annurev-conmatphys-031214-014726} \BibitemShut
  {NoStop}%
\bibitem [{\citenamefont {Alet}\ and\ \citenamefont
  {Laflorencie}(2018)}]{Alet:rev}%
  \BibitemOpen
  \bibfield  {author} {\bibinfo {author} {\bibfnamefont {Fabien}\ \bibnamefont
  {Alet}}\ and\ \bibinfo {author} {\bibfnamefont {Nicolas}\ \bibnamefont
  {Laflorencie}},\ }\bibfield  {title} {\enquote {\bibinfo {title} {Many-body
  localization: An introduction and selected topics},}\ }\href {\doibase
  https://doi.org/10.1016/j.crhy.2018.03.003} {\bibfield  {journal} {\bibinfo
  {journal} {Comptes Rendus Physique}\ } (\bibinfo {year} {2018}),\
  https://doi.org/10.1016/j.crhy.2018.03.003}\BibitemShut {NoStop}%
\bibitem [{\citenamefont {Abanin}\ \emph {et~al.}(2019)\citenamefont {Abanin},
  \citenamefont {Altman}, \citenamefont {Bloch},\ and\ \citenamefont
  {Serbyn}}]{Abanin:rev}%
  \BibitemOpen
  \bibfield  {author} {\bibinfo {author} {\bibfnamefont {Dmitry~A.}\
  \bibnamefont {Abanin}}, \bibinfo {author} {\bibfnamefont {Ehud}\ \bibnamefont
  {Altman}}, \bibinfo {author} {\bibfnamefont {Immanuel}\ \bibnamefont
  {Bloch}}, \ and\ \bibinfo {author} {\bibfnamefont {Maksym}\ \bibnamefont
  {Serbyn}},\ }\bibfield  {title} {\enquote {\bibinfo {title} {Colloquium:
  Many-body localization, thermalization, and entanglement},}\ }\href {\doibase
  10.1103/RevModPhys.91.021001} {\bibfield  {journal} {\bibinfo  {journal}
  {Rev. Mod. Phys.}\ }\textbf {\bibinfo {volume} {91}},\ \bibinfo {pages}
  {021001} (\bibinfo {year} {2019})}\BibitemShut {NoStop}%
\bibitem [{\citenamefont {Imbrie}(2016)}]{imbrie_diagonalization_2016}%
  \BibitemOpen
  \bibfield  {author} {\bibinfo {author} {\bibfnamefont {John~Z.}\ \bibnamefont
  {Imbrie}},\ }\bibfield  {title} {\enquote {\bibinfo {title} {Diagonalization
  and {Many}-{Body} {Localization} for a {Disordered} {Quantum} {Spin}
  {Chain}},}\ }\href {\doibase 10.1103/PhysRevLett.117.027201} {\bibfield
  {journal} {\bibinfo  {journal} {Phys. Rev. Lett.}\ }\textbf {\bibinfo
  {volume} {117}},\ \bibinfo {pages} {027201} (\bibinfo {year}
  {2016})}\BibitemShut {NoStop}%
\bibitem [{\citenamefont {D'Alessio}\ and\ \citenamefont
  {Polkovnikov}(2013)}]{Alessio:2013}%
  \BibitemOpen
  \bibfield  {author} {\bibinfo {author} {\bibfnamefont {Luca}\ \bibnamefont
  {D'Alessio}}\ and\ \bibinfo {author} {\bibfnamefont {Anatoli}\ \bibnamefont
  {Polkovnikov}},\ }\bibfield  {title} {\enquote {\bibinfo {title} {Many-body
  energy localization transition in periodically driven systems},}\ }\href
  {\doibase https://doi.org/10.1016/j.aop.2013.02.011} {\bibfield  {journal}
  {\bibinfo  {journal} {Annals of Physics}\ }\textbf {\bibinfo {volume}
  {333}},\ \bibinfo {pages} {19 -- 33} (\bibinfo {year} {2013})}\BibitemShut
  {NoStop}%
\bibitem [{\citenamefont {Lazarides}\ \emph {et~al.}(2015)\citenamefont
  {Lazarides}, \citenamefont {Das},\ and\ \citenamefont
  {Moessner}}]{Lazarides:2015}%
  \BibitemOpen
  \bibfield  {author} {\bibinfo {author} {\bibfnamefont {Achilleas}\
  \bibnamefont {Lazarides}}, \bibinfo {author} {\bibfnamefont {Arnab}\
  \bibnamefont {Das}}, \ and\ \bibinfo {author} {\bibfnamefont {Roderich}\
  \bibnamefont {Moessner}},\ }\bibfield  {title} {\enquote {\bibinfo {title}
  {Fate of many-body localization under periodic driving},}\ }\href {\doibase
  10.1103/PhysRevLett.115.030402} {\bibfield  {journal} {\bibinfo  {journal}
  {Phys. Rev. Lett.}\ }\textbf {\bibinfo {volume} {115}},\ \bibinfo {pages}
  {030402} (\bibinfo {year} {2015})}\BibitemShut {NoStop}%
\bibitem [{\citenamefont {Ponte}\ \emph
  {et~al.}(2015{\natexlab{a}})\citenamefont {Ponte}, \citenamefont {Chandran},
  \citenamefont {Papi{\'c}},\ and\ \citenamefont {Abanin}}]{Ponte:2015}%
  \BibitemOpen
  \bibfield  {author} {\bibinfo {author} {\bibfnamefont {Pedro}\ \bibnamefont
  {Ponte}}, \bibinfo {author} {\bibfnamefont {Anushya}\ \bibnamefont
  {Chandran}}, \bibinfo {author} {\bibfnamefont {Z.}~\bibnamefont {Papi{\'c}}},
  \ and\ \bibinfo {author} {\bibfnamefont {Dmitry~A.}\ \bibnamefont {Abanin}},\
  }\bibfield  {title} {\enquote {\bibinfo {title} {Periodically driven ergodic
  and many-body localized quantum systems},}\ }\href {\doibase
  https://doi.org/10.1016/j.aop.2014.11.008} {\bibfield  {journal} {\bibinfo
  {journal} {Annals of Physics}\ }\textbf {\bibinfo {volume} {353}},\ \bibinfo
  {pages} {196 -- 204} (\bibinfo {year} {2015}{\natexlab{a}})}\BibitemShut
  {NoStop}%
\bibitem [{\citenamefont {Ponte}\ \emph
  {et~al.}(2015{\natexlab{b}})\citenamefont {Ponte}, \citenamefont
  {Papi\ifmmode~\acute{c}\else \'{c}\fi{}}, \citenamefont {Huveneers},\ and\
  \citenamefont {Abanin}}]{Ponte:2015prl}%
  \BibitemOpen
  \bibfield  {author} {\bibinfo {author} {\bibfnamefont {Pedro}\ \bibnamefont
  {Ponte}}, \bibinfo {author} {\bibfnamefont {Z.}~\bibnamefont
  {Papi\ifmmode~\acute{c}\else \'{c}\fi{}}}, \bibinfo {author} {\bibfnamefont
  {François}\ \bibnamefont {Huveneers}}, \ and\ \bibinfo {author}
  {\bibfnamefont {Dmitry~A.}\ \bibnamefont {Abanin}},\ }\bibfield  {title}
  {\enquote {\bibinfo {title} {Many-body localization in periodically driven
  systems},}\ }\href {\doibase 10.1103/PhysRevLett.114.140401} {\bibfield
  {journal} {\bibinfo  {journal} {Phys. Rev. Lett.}\ }\textbf {\bibinfo
  {volume} {114}},\ \bibinfo {pages} {140401} (\bibinfo {year}
  {2015}{\natexlab{b}})}\BibitemShut {NoStop}%
\bibitem [{\citenamefont {Abanin}\ \emph {et~al.}(2016)\citenamefont {Abanin},
  \citenamefont {De~Roeck},\ and\ \citenamefont {Huveneers}}]{Abanin:2016}%
  \BibitemOpen
  \bibfield  {author} {\bibinfo {author} {\bibfnamefont {Dmitry~A.}\
  \bibnamefont {Abanin}}, \bibinfo {author} {\bibfnamefont {Wojciech}\
  \bibnamefont {De~Roeck}}, \ and\ \bibinfo {author} {\bibfnamefont
  {François}\ \bibnamefont {Huveneers}},\ }\bibfield  {title} {\enquote
  {\bibinfo {title} {Theory of many-body localization in periodically driven
  systems},}\ }\href {\doibase https://doi.org/10.1016/j.aop.2016.03.010}
  {\bibfield  {journal} {\bibinfo  {journal} {Annals of Physics}\ }\textbf
  {\bibinfo {volume} {372}},\ \bibinfo {pages} {1 -- 11} (\bibinfo {year}
  {2016})}\BibitemShut {NoStop}%
\bibitem [{\citenamefont {Schreiber}\ \emph {et~al.}(2015)\citenamefont
  {Schreiber}, \citenamefont {Hodgman}, \citenamefont {Bordia}, \citenamefont
  {L{\"u}schen}, \citenamefont {Fischer}, \citenamefont {Vosk}, \citenamefont
  {Altman}, \citenamefont {Schneider},\ and\ \citenamefont
  {Bloch}}]{Schreiber:2015}%
  \BibitemOpen
  \bibfield  {author} {\bibinfo {author} {\bibfnamefont {Michael}\ \bibnamefont
  {Schreiber}}, \bibinfo {author} {\bibfnamefont {Sean~S}\ \bibnamefont
  {Hodgman}}, \bibinfo {author} {\bibfnamefont {Pranjal}\ \bibnamefont
  {Bordia}}, \bibinfo {author} {\bibfnamefont {Henrik~P}\ \bibnamefont
  {L{\"u}schen}}, \bibinfo {author} {\bibfnamefont {Mark~H}\ \bibnamefont
  {Fischer}}, \bibinfo {author} {\bibfnamefont {Ronen}\ \bibnamefont {Vosk}},
  \bibinfo {author} {\bibfnamefont {Ehud}\ \bibnamefont {Altman}}, \bibinfo
  {author} {\bibfnamefont {Ulrich}\ \bibnamefont {Schneider}}, \ and\ \bibinfo
  {author} {\bibfnamefont {Immanuel}\ \bibnamefont {Bloch}},\ }\bibfield
  {title} {\enquote {\bibinfo {title} {{Observation of many-body localization
  of interacting fermions in a quasirandom optical lattice}},}\ }\href
  {http://www.sciencemag.org/cgi/doi/10.1126/science.aaa7432} {\bibfield
  {journal} {\bibinfo  {journal} {Science}\ }\textbf {\bibinfo {volume}
  {349}},\ \bibinfo {pages} {842--845} (\bibinfo {year} {2015})}\BibitemShut
  {NoStop}%
\bibitem [{\citenamefont {Smith}\ \emph {et~al.}(2016)\citenamefont {Smith},
  \citenamefont {Lee}, \citenamefont {Richerme}, \citenamefont {Neyenhuis},
  \citenamefont {Hess}, \citenamefont {Hauke}, \citenamefont {Heyl},
  \citenamefont {Huse},\ and\ \citenamefont {Monroe}}]{Smith:2016}%
  \BibitemOpen
  \bibfield  {author} {\bibinfo {author} {\bibfnamefont {J}~\bibnamefont
  {Smith}}, \bibinfo {author} {\bibfnamefont {A}~\bibnamefont {Lee}}, \bibinfo
  {author} {\bibfnamefont {P}~\bibnamefont {Richerme}}, \bibinfo {author}
  {\bibfnamefont {B}~\bibnamefont {Neyenhuis}}, \bibinfo {author}
  {\bibfnamefont {P~W}\ \bibnamefont {Hess}}, \bibinfo {author} {\bibfnamefont
  {P}~\bibnamefont {Hauke}}, \bibinfo {author} {\bibfnamefont {M}~\bibnamefont
  {Heyl}}, \bibinfo {author} {\bibfnamefont {D~A}\ \bibnamefont {Huse}}, \ and\
  \bibinfo {author} {\bibfnamefont {C}~\bibnamefont {Monroe}},\ }\bibfield
  {title} {\enquote {\bibinfo {title} {Many-body localization in a quantum
  simulator with programmable random disorder},}\ }\href {\doibase
  10.1038/nphys3783} {\bibfield  {journal} {\bibinfo  {journal} {Nature
  Physics}\ } (\bibinfo {year} {2016}),\ 10.1038/nphys3783}\BibitemShut
  {NoStop}%
\bibitem [{\citenamefont {{\v Z}nidari{\v c}}\ \emph
  {et~al.}(2008)\citenamefont {{\v Z}nidari{\v c}}, \citenamefont {Prosen},\
  and\ \citenamefont {Prelov{\v s}ek}}]{Znidaric:2008}%
  \BibitemOpen
  \bibfield  {author} {\bibinfo {author} {\bibfnamefont {Marko}\ \bibnamefont
  {{\v Z}nidari{\v c}}}, \bibinfo {author} {\bibfnamefont {Toma{\v z}}\
  \bibnamefont {Prosen}}, \ and\ \bibinfo {author} {\bibfnamefont {Peter}\
  \bibnamefont {Prelov{\v s}ek}},\ }\bibfield  {title} {\enquote {\bibinfo
  {title} {{Many-body localization in the Heisenberg XXZ magnet in a random
  field}},}\ }\href {http://link.aps.org/doi/10.1103/PhysRevB.77.064426}
  {\bibfield  {journal} {\bibinfo  {journal} {Phys. Rev. B}\ }\textbf {\bibinfo
  {volume} {77}},\ \bibinfo {pages} {064426} (\bibinfo {year}
  {2008})}\BibitemShut {NoStop}%
\bibitem [{\citenamefont {Bardarson}\ \emph {et~al.}(2012)\citenamefont
  {Bardarson}, \citenamefont {Pollmann},\ and\ \citenamefont
  {Moore}}]{Bardarson:2012}%
  \BibitemOpen
  \bibfield  {author} {\bibinfo {author} {\bibfnamefont {Jens~H}\ \bibnamefont
  {Bardarson}}, \bibinfo {author} {\bibfnamefont {Frank}\ \bibnamefont
  {Pollmann}}, \ and\ \bibinfo {author} {\bibfnamefont {Joel~E}\ \bibnamefont
  {Moore}},\ }\bibfield  {title} {\enquote {\bibinfo {title} {{Unbounded Growth
  of Entanglement in Models of Many-Body Localization}},}\ }\href
  {http://link.aps.org/doi/10.1103/PhysRevLett.109.017202} {\bibfield
  {journal} {\bibinfo  {journal} {Phys. Rev. Lett.}\ }\textbf {\bibinfo
  {volume} {109}},\ \bibinfo {pages} {017202} (\bibinfo {year}
  {2012})}\BibitemShut {NoStop}%
\bibitem [{\citenamefont {Serbyn}\ \emph
  {et~al.}(2013{\natexlab{a}})\citenamefont {Serbyn}, \citenamefont
  {Papi{\'c}},\ and\ \citenamefont {Abanin}}]{serbyn_local_2013}%
  \BibitemOpen
  \bibfield  {author} {\bibinfo {author} {\bibfnamefont {Maksym}\ \bibnamefont
  {Serbyn}}, \bibinfo {author} {\bibfnamefont {Z.}~\bibnamefont {Papi{\'c}}}, \
  and\ \bibinfo {author} {\bibfnamefont {Dmitry~A.}\ \bibnamefont {Abanin}},\
  }\bibfield  {title} {\enquote {\bibinfo {title} {Local {Conservation} {Laws}
  and the {Structure} of the {Many}-{Body} {Localized} {States}},}\ }\href
  {\doibase 10.1103/PhysRevLett.111.127201} {\bibfield  {journal} {\bibinfo
  {journal} {Phys. Rev. Lett.}\ }\textbf {\bibinfo {volume} {111}},\ \bibinfo
  {pages} {127201} (\bibinfo {year} {2013}{\natexlab{a}})}\BibitemShut
  {NoStop}%
\bibitem [{\citenamefont {Serbyn}\ \emph
  {et~al.}(2013{\natexlab{b}})\citenamefont {Serbyn}, \citenamefont
  {Papi{\'c}},\ and\ \citenamefont {Abanin}}]{serbyn_universal_2013}%
  \BibitemOpen
  \bibfield  {author} {\bibinfo {author} {\bibfnamefont {Maksym}\ \bibnamefont
  {Serbyn}}, \bibinfo {author} {\bibfnamefont {Z.}~\bibnamefont {Papi{\'c}}}, \
  and\ \bibinfo {author} {\bibfnamefont {Dmitry~A.}\ \bibnamefont {Abanin}},\
  }\bibfield  {title} {\enquote {\bibinfo {title} {Universal {Slow} {Growth} of
  {Entanglement} in {Interacting} {Strongly} {Disordered} {Systems}},}\ }\href
  {\doibase 10.1103/PhysRevLett.110.260601} {\bibfield  {journal} {\bibinfo
  {journal} {Phys. Rev. Lett.}\ }\textbf {\bibinfo {volume} {110}},\ \bibinfo
  {pages} {260601} (\bibinfo {year} {2013}{\natexlab{b}})}\BibitemShut
  {NoStop}%
\bibitem [{\citenamefont {Huse}\ \emph {et~al.}(2014)\citenamefont {Huse},
  \citenamefont {Nandkishore},\ and\ \citenamefont
  {Oganesyan}}]{huse_phenomenology_2014}%
  \BibitemOpen
  \bibfield  {author} {\bibinfo {author} {\bibfnamefont {David~A.}\
  \bibnamefont {Huse}}, \bibinfo {author} {\bibfnamefont {Rahul}\ \bibnamefont
  {Nandkishore}}, \ and\ \bibinfo {author} {\bibfnamefont {Vadim}\ \bibnamefont
  {Oganesyan}},\ }\bibfield  {title} {\enquote {\bibinfo {title} {Phenomenology
  of fully many-body-localized systems},}\ }\href {\doibase
  10.1103/PhysRevB.90.174202} {\bibfield  {journal} {\bibinfo  {journal} {Phys.
  Rev. B}\ }\textbf {\bibinfo {volume} {90}},\ \bibinfo {pages} {174202}
  (\bibinfo {year} {2014})}\BibitemShut {NoStop}%
\bibitem [{\citenamefont {Zanardi}(2001)}]{Zinardi:2001}%
  \BibitemOpen
  \bibfield  {author} {\bibinfo {author} {\bibfnamefont {Paolo}\ \bibnamefont
  {Zanardi}},\ }\bibfield  {title} {\enquote {\bibinfo {title} {Entanglement of
  quantum evolutions},}\ }\href {\doibase 10.1103/PhysRevA.63.040304}
  {\bibfield  {journal} {\bibinfo  {journal} {Phys. Rev. A}\ }\textbf {\bibinfo
  {volume} {63}},\ \bibinfo {pages} {040304} (\bibinfo {year}
  {2001})}\BibitemShut {NoStop}%
\bibitem [{\citenamefont {Prosen}\ and\ \citenamefont
  {Pi\ifmmode~\check{z}\else \v{z}\fi{}orn}(2007)}]{Prosen:2007}%
  \BibitemOpen
  \bibfield  {author} {\bibinfo {author} {\bibfnamefont {Toma{\v z}}\
  \bibnamefont {Prosen}}\ and\ \bibinfo {author} {\bibfnamefont {Iztok}\
  \bibnamefont {Pi\ifmmode~\check{z}\else \v{z}\fi{}orn}},\ }\bibfield  {title}
  {\enquote {\bibinfo {title} {Operator space entanglement entropy in a
  transverse ising chain},}\ }\href {\doibase 10.1103/PhysRevA.76.032316}
  {\bibfield  {journal} {\bibinfo  {journal} {Phys. Rev. A}\ }\textbf {\bibinfo
  {volume} {76}},\ \bibinfo {pages} {032316} (\bibinfo {year}
  {2007})}\BibitemShut {NoStop}%
\bibitem [{\citenamefont {Zhou}\ and\ \citenamefont {Luitz}(2017)}]{Zhou:2017}%
  \BibitemOpen
  \bibfield  {author} {\bibinfo {author} {\bibfnamefont {Tianci}\ \bibnamefont
  {Zhou}}\ and\ \bibinfo {author} {\bibfnamefont {David~J.}\ \bibnamefont
  {Luitz}},\ }\bibfield  {title} {\enquote {\bibinfo {title} {Operator
  entanglement entropy of the time evolution operator in chaotic systems},}\
  }\href {\doibase 10.1103/PhysRevB.95.094206} {\bibfield  {journal} {\bibinfo
  {journal} {Phys. Rev. B}\ }\textbf {\bibinfo {volume} {95}},\ \bibinfo
  {pages} {094206} (\bibinfo {year} {2017})}\BibitemShut {NoStop}%
\bibitem [{\citenamefont {Dubail}(2017)}]{Dubail:2017}%
  \BibitemOpen
  \bibfield  {author} {\bibinfo {author} {\bibfnamefont {J}~\bibnamefont
  {Dubail}},\ }\bibfield  {title} {\enquote {\bibinfo {title} {Entanglement
  scaling of operators: a conformal field theory approach, with a glimpse of
  simulability of long-time dynamics in 1+1d},}\ }\href {\doibase
  10.1088/1751-8121/aa6f38} {\bibfield  {journal} {\bibinfo  {journal} {Journal
  of Physics A: Mathematical and Theoretical}\ }\textbf {\bibinfo {volume}
  {50}},\ \bibinfo {pages} {234001} (\bibinfo {year} {2017})}\BibitemShut
  {NoStop}%
\bibitem [{\citenamefont {Pal}\ and\ \citenamefont
  {Lakshminarayan}(2018)}]{Pal:2018}%
  \BibitemOpen
  \bibfield  {author} {\bibinfo {author} {\bibfnamefont {Rajarshi}\
  \bibnamefont {Pal}}\ and\ \bibinfo {author} {\bibfnamefont {Arul}\
  \bibnamefont {Lakshminarayan}},\ }\bibfield  {title} {\enquote {\bibinfo
  {title} {Entangling power of time-evolution operators in integrable and
  nonintegrable many-body systems},}\ }\href
  {https://link.aps.org/doi/10.1103/PhysRevB.98.174304} {\bibfield  {journal}
  {\bibinfo  {journal} {Phys. Rev. B}\ }\textbf {\bibinfo {volume} {98}},\
  \bibinfo {pages} {174304} (\bibinfo {year} {2018})}\BibitemShut {NoStop}%
\bibitem [{\citenamefont {Luitz}\ and\ \citenamefont
  {Bar~Lev}(2016)}]{Luitz_anomalous:2016}%
  \BibitemOpen
  \bibfield  {author} {\bibinfo {author} {\bibfnamefont {David~J.}\
  \bibnamefont {Luitz}}\ and\ \bibinfo {author} {\bibfnamefont {Yevgeny}\
  \bibnamefont {Bar~Lev}},\ }\bibfield  {title} {\enquote {\bibinfo {title}
  {Anomalous thermalization in ergodic systems},}\ }\href {\doibase
  10.1103/PhysRevLett.117.170404} {\bibfield  {journal} {\bibinfo  {journal}
  {Phys. Rev. Lett.}\ }\textbf {\bibinfo {volume} {117}},\ \bibinfo {pages}
  {170404} (\bibinfo {year} {2016})}\BibitemShut {NoStop}%
\bibitem [{\citenamefont {Roy}\ \emph {et~al.}(2018)\citenamefont {Roy},
  \citenamefont {Bar~Lev},\ and\ \citenamefont {Luitz}}]{Roy:2018}%
  \BibitemOpen
  \bibfield  {author} {\bibinfo {author} {\bibfnamefont {Sthitadhi}\
  \bibnamefont {Roy}}, \bibinfo {author} {\bibfnamefont {Yevgeny}\ \bibnamefont
  {Bar~Lev}}, \ and\ \bibinfo {author} {\bibfnamefont {David~J.}\ \bibnamefont
  {Luitz}},\ }\bibfield  {title} {\enquote {\bibinfo {title} {Anomalous
  thermalization and transport in disordered interacting floquet systems},}\
  }\href {\doibase 10.1103/PhysRevB.98.060201} {\bibfield  {journal} {\bibinfo
  {journal} {Phys. Rev. B}\ }\textbf {\bibinfo {volume} {98}},\ \bibinfo
  {pages} {060201} (\bibinfo {year} {2018})}\BibitemShut {NoStop}%
\bibitem [{\citenamefont {Herviou}\ \emph {et~al.}(2019)\citenamefont
  {Herviou}, \citenamefont {Bera},\ and\ \citenamefont
  {Bardarson}}]{Herviou:2019}%
  \BibitemOpen
  \bibfield  {author} {\bibinfo {author} {\bibfnamefont {Lo\"{\i}c}\
  \bibnamefont {Herviou}}, \bibinfo {author} {\bibfnamefont {Soumya}\
  \bibnamefont {Bera}}, \ and\ \bibinfo {author} {\bibfnamefont {Jens~H.}\
  \bibnamefont {Bardarson}},\ }\bibfield  {title} {\enquote {\bibinfo {title}
  {Multiscale entanglement clusters at the many-body localization phase
  transition},}\ }\href {\doibase 10.1103/PhysRevB.99.134205} {\bibfield
  {journal} {\bibinfo  {journal} {Phys. Rev. B}\ }\textbf {\bibinfo {volume}
  {99}},\ \bibinfo {pages} {134205} (\bibinfo {year} {2019})}\BibitemShut
  {NoStop}%
\bibitem [{\citenamefont {B\"acker}\ \emph {et~al.}(2019)\citenamefont
  {B\"acker}, \citenamefont {Haque},\ and\ \citenamefont
  {Khaymovich}}]{Baecker:2019}%
  \BibitemOpen
  \bibfield  {author} {\bibinfo {author} {\bibfnamefont {Arnd}\ \bibnamefont
  {B\"acker}}, \bibinfo {author} {\bibfnamefont {Masudul}\ \bibnamefont
  {Haque}}, \ and\ \bibinfo {author} {\bibfnamefont {Ivan~M.}\ \bibnamefont
  {Khaymovich}},\ }\bibfield  {title} {\enquote {\bibinfo {title} {Multifractal
  dimensions for random matrices, chaotic quantum maps, and many-body
  systems},}\ }\href {\doibase 10.1103/PhysRevE.100.032117} {\bibfield
  {journal} {\bibinfo  {journal} {Phys. Rev. E}\ }\textbf {\bibinfo {volume}
  {100}},\ \bibinfo {pages} {032117} (\bibinfo {year} {2019})}\BibitemShut
  {NoStop}%
\bibitem [{\citenamefont {{Colmenarez}}\ \emph {et~al.}(2019)\citenamefont
  {{Colmenarez}}, \citenamefont {{McClarty}}, \citenamefont {{Haque}},\ and\
  \citenamefont {{Luitz}}}]{Colmenarez:2019arXiv}%
  \BibitemOpen
  \bibfield  {author} {\bibinfo {author} {\bibfnamefont {Luis}\ \bibnamefont
  {{Colmenarez}}}, \bibinfo {author} {\bibfnamefont {Paul~A.}\ \bibnamefont
  {{McClarty}}}, \bibinfo {author} {\bibfnamefont {Masudul}\ \bibnamefont
  {{Haque}}}, \ and\ \bibinfo {author} {\bibfnamefont {David~J.}\ \bibnamefont
  {{Luitz}}},\ }\bibfield  {title} {\enquote {\bibinfo {title} {{Statistics of
  correlations across the many-body localization transition}},}\ }\href@noop {}
  {\bibfield  {journal} {\bibinfo  {journal} {arXiv e-prints}\ ,\ \bibinfo
  {eid} {arXiv:1906.10701}} (\bibinfo {year} {2019})},\ \Eprint
  {http://arxiv.org/abs/1906.10701} {arXiv:1906.10701 [cond-mat.dis-nn]}
  \BibitemShut {NoStop}%
\bibitem [{\citenamefont {Bar~Lev}\ and\ \citenamefont
  {Reichman}(2014)}]{BarLev:2014}%
  \BibitemOpen
  \bibfield  {author} {\bibinfo {author} {\bibfnamefont {Yevgeny}\ \bibnamefont
  {Bar~Lev}}\ and\ \bibinfo {author} {\bibfnamefont {David~R.}\ \bibnamefont
  {Reichman}},\ }\bibfield  {title} {\enquote {\bibinfo {title} {Dynamics of
  many-body localization},}\ }\href {\doibase 10.1103/PhysRevB.89.220201}
  {\bibfield  {journal} {\bibinfo  {journal} {Phys. Rev. B}\ }\textbf {\bibinfo
  {volume} {89}},\ \bibinfo {pages} {220201} (\bibinfo {year}
  {2014})}\BibitemShut {NoStop}%
\bibitem [{\citenamefont {Bar~Lev}\ \emph {et~al.}(2015)\citenamefont
  {Bar~Lev}, \citenamefont {Cohen},\ and\ \citenamefont
  {Reichman}}]{BarLev:2015}%
  \BibitemOpen
  \bibfield  {author} {\bibinfo {author} {\bibfnamefont {Yevgeny}\ \bibnamefont
  {Bar~Lev}}, \bibinfo {author} {\bibfnamefont {Guy}\ \bibnamefont {Cohen}}, \
  and\ \bibinfo {author} {\bibfnamefont {David~R.}\ \bibnamefont {Reichman}},\
  }\bibfield  {title} {\enquote {\bibinfo {title} {Absence of diffusion in an
  interacting system of spinless fermions on a one-dimensional disordered
  lattice},}\ }\href {\doibase 10.1103/PhysRevLett.114.100601} {\bibfield
  {journal} {\bibinfo  {journal} {Phys. Rev. Lett.}\ }\textbf {\bibinfo
  {volume} {114}},\ \bibinfo {pages} {100601} (\bibinfo {year}
  {2015})}\BibitemShut {NoStop}%
\bibitem [{\citenamefont {Agarwal}\ \emph {et~al.}(2015)\citenamefont
  {Agarwal}, \citenamefont {Gopalakrishnan}, \citenamefont {Knap},
  \citenamefont {M\"uller},\ and\ \citenamefont {Demler}}]{Agarwal:2015}%
  \BibitemOpen
  \bibfield  {author} {\bibinfo {author} {\bibfnamefont {Kartiek}\ \bibnamefont
  {Agarwal}}, \bibinfo {author} {\bibfnamefont {Sarang}\ \bibnamefont
  {Gopalakrishnan}}, \bibinfo {author} {\bibfnamefont {Michael}\ \bibnamefont
  {Knap}}, \bibinfo {author} {\bibfnamefont {Markus}\ \bibnamefont {M\"uller}},
  \ and\ \bibinfo {author} {\bibfnamefont {Eugene}\ \bibnamefont {Demler}},\
  }\bibfield  {title} {\enquote {\bibinfo {title} {Anomalous diffusion and
  griffiths effects near the many-body localization transition},}\ }\href
  {\doibase 10.1103/PhysRevLett.114.160401} {\bibfield  {journal} {\bibinfo
  {journal} {Phys. Rev. Lett.}\ }\textbf {\bibinfo {volume} {114}},\ \bibinfo
  {pages} {160401} (\bibinfo {year} {2015})}\BibitemShut {NoStop}%
\bibitem [{\citenamefont {Potter}\ \emph {et~al.}(2015)\citenamefont {Potter},
  \citenamefont {Vasseur},\ and\ \citenamefont {Parameswaran}}]{Potter:2015}%
  \BibitemOpen
  \bibfield  {author} {\bibinfo {author} {\bibfnamefont {Andrew~C.}\
  \bibnamefont {Potter}}, \bibinfo {author} {\bibfnamefont {Romain}\
  \bibnamefont {Vasseur}}, \ and\ \bibinfo {author} {\bibfnamefont {S.~A.}\
  \bibnamefont {Parameswaran}},\ }\bibfield  {title} {\enquote {\bibinfo
  {title} {Universal properties of many-body delocalization transitions},}\
  }\href {\doibase 10.1103/PhysRevX.5.031033} {\bibfield  {journal} {\bibinfo
  {journal} {Phys. Rev. X}\ }\textbf {\bibinfo {volume} {5}},\ \bibinfo {pages}
  {031033} (\bibinfo {year} {2015})}\BibitemShut {NoStop}%
\bibitem [{\citenamefont {Vosk}\ \emph {et~al.}(2015)\citenamefont {Vosk},
  \citenamefont {Huse},\ and\ \citenamefont {Altman}}]{Vosk:2015}%
  \BibitemOpen
  \bibfield  {author} {\bibinfo {author} {\bibfnamefont {Ronen}\ \bibnamefont
  {Vosk}}, \bibinfo {author} {\bibfnamefont {David~A.}\ \bibnamefont {Huse}}, \
  and\ \bibinfo {author} {\bibfnamefont {Ehud}\ \bibnamefont {Altman}},\
  }\bibfield  {title} {\enquote {\bibinfo {title} {Theory of the many-body
  localization transition in one-dimensional systems},}\ }\href {\doibase
  10.1103/PhysRevX.5.031032} {\bibfield  {journal} {\bibinfo  {journal} {Phys.
  Rev. X}\ }\textbf {\bibinfo {volume} {5}},\ \bibinfo {pages} {031032}
  (\bibinfo {year} {2015})}\BibitemShut {NoStop}%
\bibitem [{\citenamefont {\ifmmode \check{Z}\else
  \v{Z}\fi{}nidari\ifmmode~\check{c}\else \v{c}\fi{}}\ \emph
  {et~al.}(2016)\citenamefont {\ifmmode \check{Z}\else
  \v{Z}\fi{}nidari\ifmmode~\check{c}\else \v{c}\fi{}}, \citenamefont
  {Scardicchio},\ and\ \citenamefont {Varma}}]{Znidaric:2016}%
  \BibitemOpen
  \bibfield  {author} {\bibinfo {author} {\bibfnamefont {Marko}\ \bibnamefont
  {\ifmmode \check{Z}\else \v{Z}\fi{}nidari\ifmmode~\check{c}\else
  \v{c}\fi{}}}, \bibinfo {author} {\bibfnamefont {Antonello}\ \bibnamefont
  {Scardicchio}}, \ and\ \bibinfo {author} {\bibfnamefont {Vipin~Kerala}\
  \bibnamefont {Varma}},\ }\bibfield  {title} {\enquote {\bibinfo {title}
  {Diffusive and subdiffusive spin transport in the ergodic phase of a
  many-body localizable system},}\ }\href {\doibase
  10.1103/PhysRevLett.117.040601} {\bibfield  {journal} {\bibinfo  {journal}
  {Phys. Rev. Lett.}\ }\textbf {\bibinfo {volume} {117}},\ \bibinfo {pages}
  {040601} (\bibinfo {year} {2016})}\BibitemShut {NoStop}%
\bibitem [{\citenamefont {Rehn}\ \emph {et~al.}(2016)\citenamefont {Rehn},
  \citenamefont {Lazarides}, \citenamefont {Pollmann},\ and\ \citenamefont
  {Moessner}}]{Rehn:2016}%
  \BibitemOpen
  \bibfield  {author} {\bibinfo {author} {\bibfnamefont {Jorge}\ \bibnamefont
  {Rehn}}, \bibinfo {author} {\bibfnamefont {Achilleas}\ \bibnamefont
  {Lazarides}}, \bibinfo {author} {\bibfnamefont {Frank}\ \bibnamefont
  {Pollmann}}, \ and\ \bibinfo {author} {\bibfnamefont {Roderich}\ \bibnamefont
  {Moessner}},\ }\bibfield  {title} {\enquote {\bibinfo {title} {How periodic
  driving heats a disordered quantum spin chain},}\ }\href {\doibase
  10.1103/PhysRevB.94.020201} {\bibfield  {journal} {\bibinfo  {journal} {Phys.
  Rev. B}\ }\textbf {\bibinfo {volume} {94}},\ \bibinfo {pages} {020201}
  (\bibinfo {year} {2016})}\BibitemShut {NoStop}%
\bibitem [{\citenamefont {Bar~Lev}\ \emph {et~al.}(2017)\citenamefont
  {Bar~Lev}, \citenamefont {Kennes}, \citenamefont {Klöckner}, \citenamefont
  {Reichman},\ and\ \citenamefont {Karrasch}}]{BarLev:2017}%
  \BibitemOpen
  \bibfield  {author} {\bibinfo {author} {\bibfnamefont {Yevgeny}\ \bibnamefont
  {Bar~Lev}}, \bibinfo {author} {\bibfnamefont {Dante~M.}\ \bibnamefont
  {Kennes}}, \bibinfo {author} {\bibfnamefont {Christian}\ \bibnamefont
  {Klöckner}}, \bibinfo {author} {\bibfnamefont {David~R.}\ \bibnamefont
  {Reichman}}, \ and\ \bibinfo {author} {\bibfnamefont {Christoph}\
  \bibnamefont {Karrasch}},\ }\bibfield  {title} {\enquote {\bibinfo {title}
  {{Transport in quasiperiodic interacting systems: From superdiffusion to
  subdiffusion}},}\ }\href {\doibase 10.1209/0295-5075/119/37003} {\bibfield
  {journal} {\bibinfo  {journal} {{EPL} (Europhysics Letters)}\ }\textbf
  {\bibinfo {volume} {119}},\ \bibinfo {pages} {37003} (\bibinfo {year}
  {2017})}\BibitemShut {NoStop}%
\bibitem [{\citenamefont {Luitz}\ \emph {et~al.}(2017)\citenamefont {Luitz},
  \citenamefont {Bar~Lev},\ and\ \citenamefont
  {Lazarides}}]{Luitz_absence:2017}%
  \BibitemOpen
  \bibfield  {author} {\bibinfo {author} {\bibfnamefont {David~J.}\
  \bibnamefont {Luitz}}, \bibinfo {author} {\bibfnamefont {Yevgeny}\
  \bibnamefont {Bar~Lev}}, \ and\ \bibinfo {author} {\bibfnamefont {Achilleas}\
  \bibnamefont {Lazarides}},\ }\bibfield  {title} {\enquote {\bibinfo {title}
  {{Absence of dynamical localization in interacting driven systems}},}\ }\href
  {\doibase 10.21468/SciPostPhys.3.4.029} {\bibfield  {journal} {\bibinfo
  {journal} {SciPost Phys.}\ }\textbf {\bibinfo {volume} {3}},\ \bibinfo
  {pages} {029} (\bibinfo {year} {2017})}\BibitemShut {NoStop}%
\bibitem [{\citenamefont {Bera}\ \emph {et~al.}(2017)\citenamefont {Bera},
  \citenamefont {De~Tomasi}, \citenamefont {Weiner},\ and\ \citenamefont
  {Evers}}]{Bera:2017}%
  \BibitemOpen
  \bibfield  {author} {\bibinfo {author} {\bibfnamefont {Soumya}\ \bibnamefont
  {Bera}}, \bibinfo {author} {\bibfnamefont {Giuseppe}\ \bibnamefont
  {De~Tomasi}}, \bibinfo {author} {\bibfnamefont {Felix}\ \bibnamefont
  {Weiner}}, \ and\ \bibinfo {author} {\bibfnamefont {Ferdinand}\ \bibnamefont
  {Evers}},\ }\bibfield  {title} {\enquote {\bibinfo {title} {Density
  propagator for many-body localization: Finite-size effects, transient
  subdiffusion, and exponential decay},}\ }\href {\doibase
  10.1103/PhysRevLett.118.196801} {\bibfield  {journal} {\bibinfo  {journal}
  {Phys. Rev. Lett.}\ }\textbf {\bibinfo {volume} {118}},\ \bibinfo {pages}
  {196801} (\bibinfo {year} {2017})}\BibitemShut {NoStop}%
\bibitem [{\citenamefont {Luitz}\ and\ \citenamefont
  {Bar~Lev}(2017{\natexlab{a}})}]{Luitz_rev:2017}%
  \BibitemOpen
  \bibfield  {author} {\bibinfo {author} {\bibfnamefont {David~J.}\
  \bibnamefont {Luitz}}\ and\ \bibinfo {author} {\bibfnamefont {Yevgeny}\
  \bibnamefont {Bar~Lev}},\ }\bibfield  {title} {\enquote {\bibinfo {title}
  {The ergodic side of the many-body localization transition},}\ }\href
  {\doibase 10.1002/andp.201600350} {\bibfield  {journal} {\bibinfo  {journal}
  {Annalen der Physik}\ }\textbf {\bibinfo {volume} {529}},\ \bibinfo {pages}
  {1600350--n/a} (\bibinfo {year} {2017}{\natexlab{a}})},\ \bibinfo {note}
  {1600350}\BibitemShut {NoStop}%
\bibitem [{\citenamefont {Agarwal}\ \emph {et~al.}(2017)\citenamefont
  {Agarwal}, \citenamefont {Altman}, \citenamefont {Demler}, \citenamefont
  {Gopalakrishnan}, \citenamefont {Huse},\ and\ \citenamefont
  {Knap}}]{Agarwal_rev:2017}%
  \BibitemOpen
  \bibfield  {author} {\bibinfo {author} {\bibfnamefont {Kartiek}\ \bibnamefont
  {Agarwal}}, \bibinfo {author} {\bibfnamefont {Ehud}\ \bibnamefont {Altman}},
  \bibinfo {author} {\bibfnamefont {Eugene}\ \bibnamefont {Demler}}, \bibinfo
  {author} {\bibfnamefont {Sarang}\ \bibnamefont {Gopalakrishnan}}, \bibinfo
  {author} {\bibfnamefont {David~A.}\ \bibnamefont {Huse}}, \ and\ \bibinfo
  {author} {\bibfnamefont {Michael}\ \bibnamefont {Knap}},\ }\bibfield  {title}
  {\enquote {\bibinfo {title} {Rare-region effects and dynamics near the
  many-body localization transition},}\ }\href {\doibase
  10.1002/andp.201600326} {\bibfield  {journal} {\bibinfo  {journal} {Annalen
  der Physik}\ }\textbf {\bibinfo {volume} {529}},\ \bibinfo {pages}
  {1600326--n/a} (\bibinfo {year} {2017})},\ \bibinfo {note}
  {1600326}\BibitemShut {NoStop}%
\bibitem [{\citenamefont {Kozarzewski}\ \emph {et~al.}(2018)\citenamefont
  {Kozarzewski}, \citenamefont {Prelov\ifmmode~\check{s}\else \v{s}\fi{}ek},\
  and\ \citenamefont {Mierzejewski}}]{Kozarzewski:2018}%
  \BibitemOpen
  \bibfield  {author} {\bibinfo {author} {\bibfnamefont {Maciej}\ \bibnamefont
  {Kozarzewski}}, \bibinfo {author} {\bibfnamefont {Peter}\ \bibnamefont
  {Prelov\ifmmode~\check{s}\else \v{s}\fi{}ek}}, \ and\ \bibinfo {author}
  {\bibfnamefont {Marcin}\ \bibnamefont {Mierzejewski}},\ }\bibfield  {title}
  {\enquote {\bibinfo {title} {Spin subdiffusion in the disordered hubbard
  chain},}\ }\href {\doibase 10.1103/PhysRevLett.120.246602} {\bibfield
  {journal} {\bibinfo  {journal} {Phys. Rev. Lett.}\ }\textbf {\bibinfo
  {volume} {120}},\ \bibinfo {pages} {246602} (\bibinfo {year}
  {2018})}\BibitemShut {NoStop}%
\bibitem [{\citenamefont {Schulz}\ \emph {et~al.}(2018)\citenamefont {Schulz},
  \citenamefont {Taylor}, \citenamefont {Hooley},\ and\ \citenamefont
  {Scardicchio}}]{Schulz:2018}%
  \BibitemOpen
  \bibfield  {author} {\bibinfo {author} {\bibfnamefont {M.}~\bibnamefont
  {Schulz}}, \bibinfo {author} {\bibfnamefont {S.~R.}\ \bibnamefont {Taylor}},
  \bibinfo {author} {\bibfnamefont {C.~A.}\ \bibnamefont {Hooley}}, \ and\
  \bibinfo {author} {\bibfnamefont {A.}~\bibnamefont {Scardicchio}},\
  }\bibfield  {title} {\enquote {\bibinfo {title} {Energy transport in a
  disordered spin chain with broken u(1) symmetry: Diffusion, subdiffusion, and
  many-body localization},}\ }\href {\doibase 10.1103/PhysRevB.98.180201}
  {\bibfield  {journal} {\bibinfo  {journal} {Phys. Rev. B}\ }\textbf {\bibinfo
  {volume} {98}},\ \bibinfo {pages} {180201} (\bibinfo {year}
  {2018})}\BibitemShut {NoStop}%
\bibitem [{\citenamefont {Doggen}\ \emph {et~al.}(2018)\citenamefont {Doggen},
  \citenamefont {Schindler}, \citenamefont {Tikhonov}, \citenamefont {Mirlin},
  \citenamefont {Neupert}, \citenamefont {Polyakov},\ and\ \citenamefont
  {Gornyi}}]{Doggen:2018}%
  \BibitemOpen
  \bibfield  {author} {\bibinfo {author} {\bibfnamefont {Elmer V.~H.}\
  \bibnamefont {Doggen}}, \bibinfo {author} {\bibfnamefont {Frank}\
  \bibnamefont {Schindler}}, \bibinfo {author} {\bibfnamefont {Konstantin~S.}\
  \bibnamefont {Tikhonov}}, \bibinfo {author} {\bibfnamefont {Alexander~D.}\
  \bibnamefont {Mirlin}}, \bibinfo {author} {\bibfnamefont {Titus}\
  \bibnamefont {Neupert}}, \bibinfo {author} {\bibfnamefont {Dmitry~G.}\
  \bibnamefont {Polyakov}}, \ and\ \bibinfo {author} {\bibfnamefont {Igor~V.}\
  \bibnamefont {Gornyi}},\ }\bibfield  {title} {\enquote {\bibinfo {title}
  {Many-body localization and delocalization in large quantum chains},}\ }\href
  {\doibase 10.1103/PhysRevB.98.174202} {\bibfield  {journal} {\bibinfo
  {journal} {Phys. Rev. B}\ }\textbf {\bibinfo {volume} {98}},\ \bibinfo
  {pages} {174202} (\bibinfo {year} {2018})}\BibitemShut {NoStop}%
\bibitem [{\citenamefont {Luitz}\ \emph {et~al.}(2016)\citenamefont {Luitz},
  \citenamefont {Laflorencie},\ and\ \citenamefont
  {Alet}}]{Luitz_extended:2016}%
  \BibitemOpen
  \bibfield  {author} {\bibinfo {author} {\bibfnamefont {David~J.}\
  \bibnamefont {Luitz}}, \bibinfo {author} {\bibfnamefont {Nicolas}\
  \bibnamefont {Laflorencie}}, \ and\ \bibinfo {author} {\bibfnamefont
  {Fabien}\ \bibnamefont {Alet}},\ }\bibfield  {title} {\enquote {\bibinfo
  {title} {Extended slow dynamical regime close to the many-body localization
  transition},}\ }\href {\doibase 10.1103/PhysRevB.93.060201} {\bibfield
  {journal} {\bibinfo  {journal} {Phys. Rev. B}\ }\textbf {\bibinfo {volume}
  {93}},\ \bibinfo {pages} {060201} (\bibinfo {year} {2016})}\BibitemShut
  {NoStop}%
\bibitem [{\citenamefont {Lezama}\ \emph {et~al.}(2019)\citenamefont {Lezama},
  \citenamefont {Bera},\ and\ \citenamefont {Bardarson}}]{Lezama:2019}%
  \BibitemOpen
  \bibfield  {author} {\bibinfo {author} {\bibfnamefont {Tal\'{\i}a L.~M.}\
  \bibnamefont {Lezama}}, \bibinfo {author} {\bibfnamefont {Soumya}\
  \bibnamefont {Bera}}, \ and\ \bibinfo {author} {\bibfnamefont {Jens~H.}\
  \bibnamefont {Bardarson}},\ }\bibfield  {title} {\enquote {\bibinfo {title}
  {Apparent slow dynamics in the ergodic phase of a driven many-body localized
  system without extensive conserved quantities},}\ }\href {\doibase
  10.1103/PhysRevB.99.161106} {\bibfield  {journal} {\bibinfo  {journal} {Phys.
  Rev. B}\ }\textbf {\bibinfo {volume} {99}},\ \bibinfo {pages} {161106}
  (\bibinfo {year} {2019})}\BibitemShut {NoStop}%
\bibitem [{\citenamefont {Berkelbach}\ and\ \citenamefont
  {Reichman}(2010)}]{Berkelbach:2010}%
  \BibitemOpen
  \bibfield  {author} {\bibinfo {author} {\bibfnamefont {Timothy~C.}\
  \bibnamefont {Berkelbach}}\ and\ \bibinfo {author} {\bibfnamefont {David~R.}\
  \bibnamefont {Reichman}},\ }\bibfield  {title} {\enquote {\bibinfo {title}
  {Conductivity of disordered quantum lattice models at infinite temperature:
  Many-body localization},}\ }\href {\doibase 10.1103/PhysRevB.81.224429}
  {\bibfield  {journal} {\bibinfo  {journal} {Phys. Rev. B}\ }\textbf {\bibinfo
  {volume} {81}},\ \bibinfo {pages} {224429} (\bibinfo {year}
  {2010})}\BibitemShut {NoStop}%
\bibitem [{\citenamefont {Pal}\ and\ \citenamefont {Huse}(2010)}]{Pal:2010}%
  \BibitemOpen
  \bibfield  {author} {\bibinfo {author} {\bibfnamefont {Arijeet}\ \bibnamefont
  {Pal}}\ and\ \bibinfo {author} {\bibfnamefont {David~A}\ \bibnamefont
  {Huse}},\ }\bibfield  {title} {\enquote {\bibinfo {title} {{Many-body
  localization phase transition}},}\ }\href
  {http://link.aps.org/doi/10.1103/PhysRevB.82.174411} {\bibfield  {journal}
  {\bibinfo  {journal} {Phys. Rev. B}\ }\textbf {\bibinfo {volume} {82}},\
  \bibinfo {pages} {174411} (\bibinfo {year} {2010})}\BibitemShut {NoStop}%
\bibitem [{\citenamefont {Luitz}\ \emph {et~al.}(2015)\citenamefont {Luitz},
  \citenamefont {Laflorencie},\ and\ \citenamefont {Alet}}]{Luitz:2015}%
  \BibitemOpen
  \bibfield  {author} {\bibinfo {author} {\bibfnamefont {David~J}\ \bibnamefont
  {Luitz}}, \bibinfo {author} {\bibfnamefont {Nicolas}\ \bibnamefont
  {Laflorencie}}, \ and\ \bibinfo {author} {\bibfnamefont {Fabien}\
  \bibnamefont {Alet}},\ }\bibfield  {title} {\enquote {\bibinfo {title}
  {{Many-body localization edge in the random-field Heisenberg chain}},}\
  }\href {http://link.aps.org/doi/10.1103/PhysRevB.91.081103} {\bibfield
  {journal} {\bibinfo  {journal} {Phys. Rev. B}\ }\textbf {\bibinfo {volume}
  {91}},\ \bibinfo {pages} {081103(R)} (\bibinfo {year} {2015})}\BibitemShut
  {NoStop}%
\bibitem [{\citenamefont {Bera}\ \emph {et~al.}(2015)\citenamefont {Bera},
  \citenamefont {Schomerus}, \citenamefont {Heidrich-Meisner},\ and\
  \citenamefont {Bardarson}}]{Bera:2015}%
  \BibitemOpen
  \bibfield  {author} {\bibinfo {author} {\bibfnamefont {Soumya}\ \bibnamefont
  {Bera}}, \bibinfo {author} {\bibfnamefont {Henning}\ \bibnamefont
  {Schomerus}}, \bibinfo {author} {\bibfnamefont {Fabian}\ \bibnamefont
  {Heidrich-Meisner}}, \ and\ \bibinfo {author} {\bibfnamefont {Jens~H.}\
  \bibnamefont {Bardarson}},\ }\bibfield  {title} {\enquote {\bibinfo {title}
  {Many-body localization characterized from a one-particle perspective},}\
  }\href {\doibase 10.1103/PhysRevLett.115.046603} {\bibfield  {journal}
  {\bibinfo  {journal} {Phys. Rev. Lett.}\ }\textbf {\bibinfo {volume} {115}},\
  \bibinfo {pages} {046603} (\bibinfo {year} {2015})}\BibitemShut {NoStop}%
\bibitem [{\citenamefont {Nauts}\ and\ \citenamefont
  {Wyatt}(1983)}]{nauts_new_1983}%
  \BibitemOpen
  \bibfield  {author} {\bibinfo {author} {\bibfnamefont {Andr{\'e}}\
  \bibnamefont {Nauts}}\ and\ \bibinfo {author} {\bibfnamefont {Robert~E.}\
  \bibnamefont {Wyatt}},\ }\bibfield  {title} {\enquote {\bibinfo {title} {New
  {Approach} to {Many}-{State} {Quantum} {Dynamics}: {The}
  {Recursive}-{Residue}-{Generation} {Method}},}\ }\href {\doibase
  10.1103/PhysRevLett.51.2238} {\bibfield  {journal} {\bibinfo  {journal}
  {Phys. Rev. Lett.}\ }\textbf {\bibinfo {volume} {51}},\ \bibinfo {pages}
  {2238--2241} (\bibinfo {year} {1983})}\BibitemShut {NoStop}%
\bibitem [{\citenamefont {Saad}(1992)}]{Saad:1992}%
  \BibitemOpen
  \bibfield  {author} {\bibinfo {author} {\bibfnamefont {Y.}~\bibnamefont
  {Saad}},\ }\bibfield  {title} {\enquote {\bibinfo {title} {Analysis of some
  krylov subspace approximations to the matrix exponential operator},}\ }\href
  {\doibase 10.1137/0729014} {\bibfield  {journal} {\bibinfo  {journal} {SIAM
  Journal on Numerical Analysis}\ }\textbf {\bibinfo {volume} {29}},\ \bibinfo
  {pages} {209--228} (\bibinfo {year} {1992})},\ \Eprint
  {http://arxiv.org/abs/https://doi.org/10.1137/0729014}
  {https://doi.org/10.1137/0729014} \BibitemShut {NoStop}%
\bibitem [{\citenamefont {Moler}\ and\ \citenamefont
  {Van~Loan}(2003)}]{moler_nineteen_2003}%
  \BibitemOpen
  \bibfield  {author} {\bibinfo {author} {\bibfnamefont {C.}~\bibnamefont
  {Moler}}\ and\ \bibinfo {author} {\bibfnamefont {C.}~\bibnamefont
  {Van~Loan}},\ }\bibfield  {title} {\enquote {\bibinfo {title} {Nineteen
  {Dubious} {Ways} to {Compute} the {Exponential} of a {Matrix},
  {Twenty}-{Five} {Years} {Later}},}\ }\href {\doibase 10.1137/S00361445024180}
  {\bibfield  {journal} {\bibinfo  {journal} {SIAM Rev.}\ }\textbf {\bibinfo
  {volume} {45}},\ \bibinfo {pages} {3--49} (\bibinfo {year}
  {2003})}\BibitemShut {NoStop}%
\bibitem [{\citenamefont {Oganesyan}\ and\ \citenamefont
  {Huse}(2007)}]{Oganesyan:2007}%
  \BibitemOpen
  \bibfield  {author} {\bibinfo {author} {\bibfnamefont {Vadim}\ \bibnamefont
  {Oganesyan}}\ and\ \bibinfo {author} {\bibfnamefont {David~A.}\ \bibnamefont
  {Huse}},\ }\bibfield  {title} {\enquote {\bibinfo {title} {Localization of
  interacting fermions at high temperature},}\ }\href {\doibase
  10.1103/PhysRevB.75.155111} {\bibfield  {journal} {\bibinfo  {journal} {Phys.
  Rev. B}\ }\textbf {\bibinfo {volume} {75}},\ \bibinfo {pages} {155111}
  (\bibinfo {year} {2007})}\BibitemShut {NoStop}%
\bibitem [{Note1()}]{Note1}%
  \BibitemOpen
  \bibinfo {note} {Averaged over $10^5$ random GUE matrices of size
  $N=100$.}\BibitemShut {Stop}%
\bibitem [{\citenamefont {Atas}\ \emph {et~al.}(2013)\citenamefont {Atas},
  \citenamefont {Bogomolny}, \citenamefont {Giraud},\ and\ \citenamefont
  {Roux}}]{Atas:2013}%
  \BibitemOpen
  \bibfield  {author} {\bibinfo {author} {\bibfnamefont {Y.~Y.}\ \bibnamefont
  {Atas}}, \bibinfo {author} {\bibfnamefont {E.}~\bibnamefont {Bogomolny}},
  \bibinfo {author} {\bibfnamefont {O.}~\bibnamefont {Giraud}}, \ and\ \bibinfo
  {author} {\bibfnamefont {G.}~\bibnamefont {Roux}},\ }\bibfield  {title}
  {\enquote {\bibinfo {title} {Distribution of the ratio of consecutive level
  spacings in random matrix ensembles},}\ }\href {\doibase
  10.1103/PhysRevLett.110.084101} {\bibfield  {journal} {\bibinfo  {journal}
  {Phys. Rev. Lett.}\ }\textbf {\bibinfo {volume} {110}},\ \bibinfo {pages}
  {084101} (\bibinfo {year} {2013})}\BibitemShut {NoStop}%
\bibitem [{Note2()}]{Note2}%
  \BibitemOpen
  \bibinfo {note} {We note that for large matrices, CUE and GUE ensembles are
  identical, which is not the case for smaller matrices \cite
  {Alessio:2014}.}\BibitemShut {Stop}%
\bibitem [{\citenamefont {Page}(1993)}]{Page_average_1993}%
  \BibitemOpen
  \bibfield  {author} {\bibinfo {author} {\bibfnamefont {Don~N.}\ \bibnamefont
  {Page}},\ }\bibfield  {title} {\enquote {\bibinfo {title} {Average entropy of
  a subsystem},}\ }\href {\doibase 10.1103/PhysRevLett.71.1291} {\bibfield
  {journal} {\bibinfo  {journal} {Phys. Rev. Lett.}\ }\textbf {\bibinfo
  {volume} {71}},\ \bibinfo {pages} {1291--1294} (\bibinfo {year}
  {1993})}\BibitemShut {NoStop}%
\bibitem [{\citenamefont {Lieb}\ and\ \citenamefont
  {Robinson}(1972)}]{Lieb:1972}%
  \BibitemOpen
  \bibfield  {author} {\bibinfo {author} {\bibfnamefont {Elliott~H.}\
  \bibnamefont {Lieb}}\ and\ \bibinfo {author} {\bibfnamefont {Derek~W.}\
  \bibnamefont {Robinson}},\ }\bibfield  {title} {\enquote {\bibinfo {title}
  {The finite group velocity of quantum spin systems},}\ }\href {\doibase
  10.1007/BF01645779} {\bibfield  {journal} {\bibinfo  {journal}
  {Communications in Mathematical Physics}\ }\textbf {\bibinfo {volume} {28}},\
  \bibinfo {pages} {251--257} (\bibinfo {year} {1972})}\BibitemShut {NoStop}%
\bibitem [{\citenamefont {{Jonay}}\ \emph {et~al.}(2018)\citenamefont
  {{Jonay}}, \citenamefont {{Huse}},\ and\ \citenamefont
  {{Nahum}}}]{Jonay:2018}%
  \BibitemOpen
  \bibfield  {author} {\bibinfo {author} {\bibfnamefont {Cheryne}\ \bibnamefont
  {{Jonay}}}, \bibinfo {author} {\bibfnamefont {David~A.}\ \bibnamefont
  {{Huse}}}, \ and\ \bibinfo {author} {\bibfnamefont {Adam}\ \bibnamefont
  {{Nahum}}},\ }\bibfield  {title} {\enquote {\bibinfo {title} {{Coarse-grained
  dynamics of operator and state entanglement}},}\ }\href@noop {} {\bibfield
  {journal} {\bibinfo  {journal} {arXiv e-prints}\ ,\ \bibinfo {eid}
  {arXiv:1803.00089}} (\bibinfo {year} {2018})},\ \Eprint
  {http://arxiv.org/abs/1803.00089} {arXiv:1803.00089 [cond-mat.stat-mech]}
  \BibitemShut {NoStop}%
\bibitem [{\citenamefont {Nanduri}\ \emph {et~al.}(2014)\citenamefont
  {Nanduri}, \citenamefont {Kim},\ and\ \citenamefont {Huse}}]{Nanduri:2014}%
  \BibitemOpen
  \bibfield  {author} {\bibinfo {author} {\bibfnamefont {Arun}\ \bibnamefont
  {Nanduri}}, \bibinfo {author} {\bibfnamefont {Hyungwon}\ \bibnamefont {Kim}},
  \ and\ \bibinfo {author} {\bibfnamefont {David~A.}\ \bibnamefont {Huse}},\
  }\bibfield  {title} {\enquote {\bibinfo {title} {Entanglement spreading in a
  many-body localized system},}\ }\href {\doibase 10.1103/PhysRevB.90.064201}
  {\bibfield  {journal} {\bibinfo  {journal} {Phys. Rev. B}\ }\textbf {\bibinfo
  {volume} {90}},\ \bibinfo {pages} {064201} (\bibinfo {year}
  {2014})}\BibitemShut {NoStop}%
\bibitem [{\citenamefont {Coffman}\ \emph {et~al.}(2000)\citenamefont
  {Coffman}, \citenamefont {Kundu},\ and\ \citenamefont
  {Wootters}}]{Coffman:2000}%
  \BibitemOpen
  \bibfield  {author} {\bibinfo {author} {\bibfnamefont {Valerie}\ \bibnamefont
  {Coffman}}, \bibinfo {author} {\bibfnamefont {Joydip}\ \bibnamefont {Kundu}},
  \ and\ \bibinfo {author} {\bibfnamefont {William~K.}\ \bibnamefont
  {Wootters}},\ }\bibfield  {title} {\enquote {\bibinfo {title} {Distributed
  entanglement},}\ }\href {\doibase 10.1103/PhysRevA.61.052306} {\bibfield
  {journal} {\bibinfo  {journal} {Phys. Rev. A}\ }\textbf {\bibinfo {volume}
  {61}},\ \bibinfo {pages} {052306} (\bibinfo {year} {2000})}\BibitemShut
  {NoStop}%
\bibitem [{\citenamefont {Osborne}\ and\ \citenamefont
  {Verstraete}(2006)}]{Osborne:2006}%
  \BibitemOpen
  \bibfield  {author} {\bibinfo {author} {\bibfnamefont {Tobias~J.}\
  \bibnamefont {Osborne}}\ and\ \bibinfo {author} {\bibfnamefont {Frank}\
  \bibnamefont {Verstraete}},\ }\bibfield  {title} {\enquote {\bibinfo {title}
  {General monogamy inequality for bipartite qubit entanglement},}\ }\href
  {\doibase 10.1103/PhysRevLett.96.220503} {\bibfield  {journal} {\bibinfo
  {journal} {Phys. Rev. Lett.}\ }\textbf {\bibinfo {volume} {96}},\ \bibinfo
  {pages} {220503} (\bibinfo {year} {2006})}\BibitemShut {NoStop}%
\bibitem [{\citenamefont {Nahum}\ \emph {et~al.}(2017)\citenamefont {Nahum},
  \citenamefont {Ruhman}, \citenamefont {Vijay},\ and\ \citenamefont
  {Haah}}]{Nahum:2017}%
  \BibitemOpen
  \bibfield  {author} {\bibinfo {author} {\bibfnamefont {Adam}\ \bibnamefont
  {Nahum}}, \bibinfo {author} {\bibfnamefont {Jonathan}\ \bibnamefont
  {Ruhman}}, \bibinfo {author} {\bibfnamefont {Sagar}\ \bibnamefont {Vijay}}, \
  and\ \bibinfo {author} {\bibfnamefont {Jeongwan}\ \bibnamefont {Haah}},\
  }\bibfield  {title} {\enquote {\bibinfo {title} {Quantum entanglement growth
  under random unitary dynamics},}\ }\href {\doibase 10.1103/PhysRevX.7.031016}
  {\bibfield  {journal} {\bibinfo  {journal} {Phys. Rev. X}\ }\textbf {\bibinfo
  {volume} {7}},\ \bibinfo {pages} {031016} (\bibinfo {year}
  {2017})}\BibitemShut {NoStop}%
\bibitem [{\citenamefont {Nahum}\ \emph {et~al.}(2018)\citenamefont {Nahum},
  \citenamefont {Ruhman},\ and\ \citenamefont {Huse}}]{Nahum:2018}%
  \BibitemOpen
  \bibfield  {author} {\bibinfo {author} {\bibfnamefont {Adam}\ \bibnamefont
  {Nahum}}, \bibinfo {author} {\bibfnamefont {Jonathan}\ \bibnamefont
  {Ruhman}}, \ and\ \bibinfo {author} {\bibfnamefont {David~A.}\ \bibnamefont
  {Huse}},\ }\bibfield  {title} {\enquote {\bibinfo {title} {Dynamics of
  entanglement and transport in one-dimensional systems with quenched
  randomness},}\ }\href {\doibase 10.1103/PhysRevB.98.035118} {\bibfield
  {journal} {\bibinfo  {journal} {Phys. Rev. B}\ }\textbf {\bibinfo {volume}
  {98}},\ \bibinfo {pages} {035118} (\bibinfo {year} {2018})}\BibitemShut
  {NoStop}%
\bibitem [{\citenamefont {Luitz}\ and\ \citenamefont
  {Bar~Lev}(2017{\natexlab{b}})}]{Luitz_information:2017}%
  \BibitemOpen
  \bibfield  {author} {\bibinfo {author} {\bibfnamefont {David~J.}\
  \bibnamefont {Luitz}}\ and\ \bibinfo {author} {\bibfnamefont {Yevgeny}\
  \bibnamefont {Bar~Lev}},\ }\bibfield  {title} {\enquote {\bibinfo {title}
  {Information propagation in isolated quantum systems},}\ }\href {\doibase
  10.1103/PhysRevB.96.020406} {\bibfield  {journal} {\bibinfo  {journal} {Phys.
  Rev. B}\ }\textbf {\bibinfo {volume} {96}},\ \bibinfo {pages} {020406}
  (\bibinfo {year} {2017}{\natexlab{b}})}\BibitemShut {NoStop}%
\end{thebibliography}%

\end{document}